\documentclass[letter,scriptaddress,twocolumn, prl,showkeys,showpacs,groupedaddress]{revtex4-1}

	\usepackage{amsmath}
	\usepackage{makeidx}
	\usepackage{amsfonts}
	\usepackage{graphicx, xcolor}  
	\usepackage{dcolumn}   
	\usepackage{bm}        
	\usepackage{amssymb}   
	\usepackage[ansinew]{inputenc}
	\usepackage[usenames,dvipsnames]{pstricks}
	\usepackage{subfigure}
	\usepackage{epstopdf}
	\usepackage{pst-grad} 
	\usepackage{pst-plot} 
	\usepackage{lineno}

	\usepackage[colorlinks,hyperindex]{hyperref}
	\hypersetup
	{
		colorlinks,%
		citecolor=black,%
		linkcolor=black,%
		urlcolor=black,%
	}

\newcommand{\be}{\begin{equation}}
\newcommand{\ee}{\end{equation}}
\newcommand{\baa}{\begin{align}}
\newcommand{\eaa}{\end{align}}

\newcommand{\tr}{\mathop{\mathrm{tr}}}

\newcommand{\mf}{\mathcal{F}}
\newcommand{\mh}{\mathcal{H}}
\newcommand{\mg}{\mathcal{G}}
\newcommand{\mk}{\mathcal{K}}
\newcommand{\ms}{\mathcal{S}}

\newcommand{\mi}{\mathcal{I}}

\newcommand{\mm}{\mathcal{M}}

\newcommand{\mz}{\mathcal{Z}}

\newcommand{\rd}{{\rm d}}

\newcommand{\lp}{\left(}
\newcommand{\rp}{\right)}

\begin{document}

\title{Entropy favors heterogeneous structures of networks near the rigidity threshold}
\author{Le Yan}
\email{lyan@kitp.ucsb.edu}
\affiliation{Kavli Institute for Theoretical Physics, University of California, Santa Barbara, CA 93106, USA}

\date{\today}

\begin{abstract}

The dynamical properties and mechanical functions of amorphous materials are governed by their microscopic structures, particularly the elasticity of the interaction networks, 
which is generally complicated by structural heterogeneity. 
This ubiquitous heterogeneous nature of amorphous materials is intriguingly attributed to a complex role of entropy. 
Here, we show in disordered networks that the vibrational entropy increases by creating phase-separated structures when the interaction connectivity is close to the onset of network rigidity. 
The stress energy, which conversely penalizes the heterogeneity, finally dominates a smaller vicinity of the rigidity threshold at the glass transition and creates a homogeneous intermediate phase.
This picture of structures changing between homogeneous and heterogeneous phases by varying connectivity provides an interpretation of the transitions observed in chalcogenide glasses. 
\end{abstract}

\maketitle
%
Lacking long-range order, amorphous materials 
are fully governed by their microscopic structures. 
Increasing evidence indicates that the structural elasticity of such materials correlates with their dynamical properties and mechanical functions, such as the suddenly slowing relaxations of glasses~\cite{Hall03, Shintani08, Mauro09, Yan13} and the allosteric regulation of proteins~\cite{Yan17,Rocks17}. 
A crucial factor behind the structural disorder that controls the linear elasticity of a structure is the average  {number of constraints $n$} of its interaction network and the rigidity transition associated with tuning {$n$}~\cite{Liu10}. 
At the Maxwell point {$n_c=d$}~\cite{Maxwell64}, which is the minimum number of constraints per particle to avoid floppy modes in spatial dimension $d$, both the elastic moduli and self-stresses vanish, accompanied by a vanishing onset frequency $\omega^*$ of the soft vibrations on the so-called boson peak~\cite{OHern03,Silbert05,Wyart05b}. 
However, it is questionable whether these results obtained in homogeneous networks apply to heterogeneous network structures, which may be fundamental.

\begin{figure}[h!]
\centering
\includegraphics[width=1.0\columnwidth]{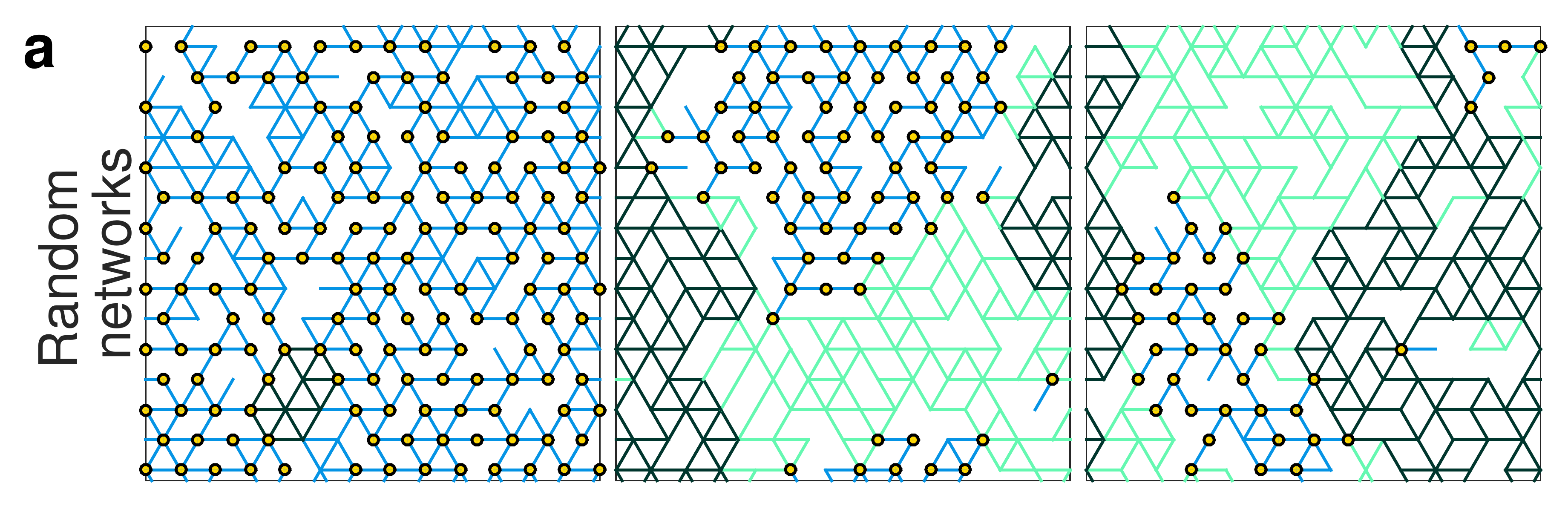}\\
\includegraphics[width=1.0\columnwidth]{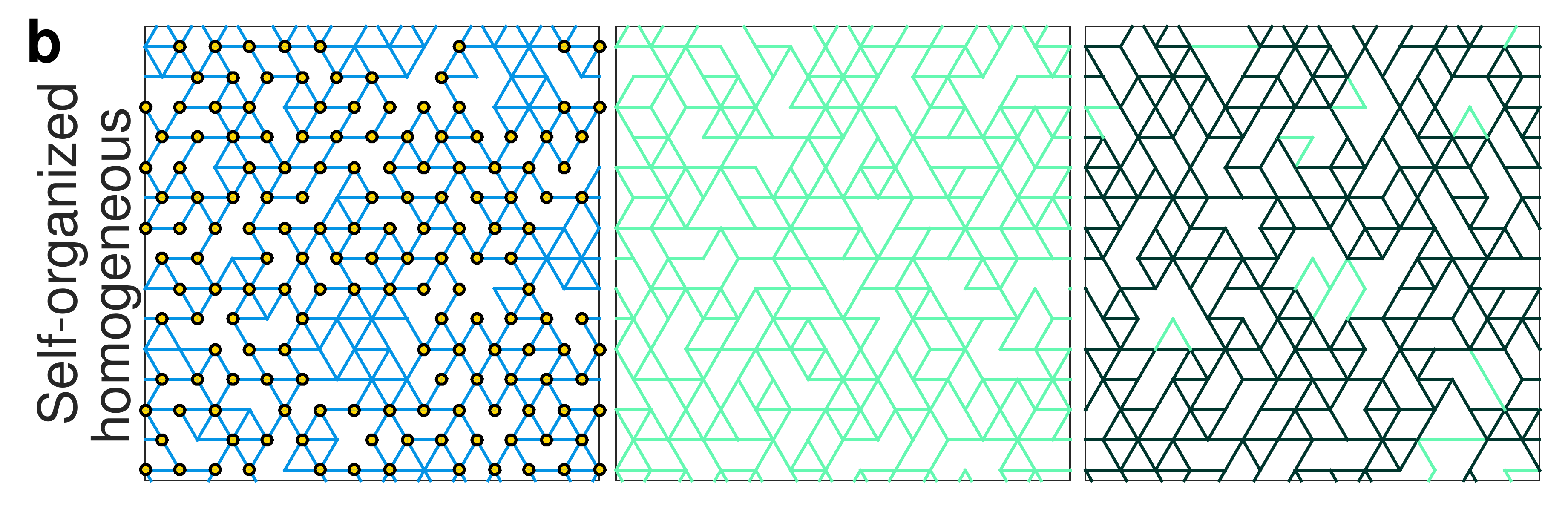}\\
\includegraphics[width=1.0\columnwidth]{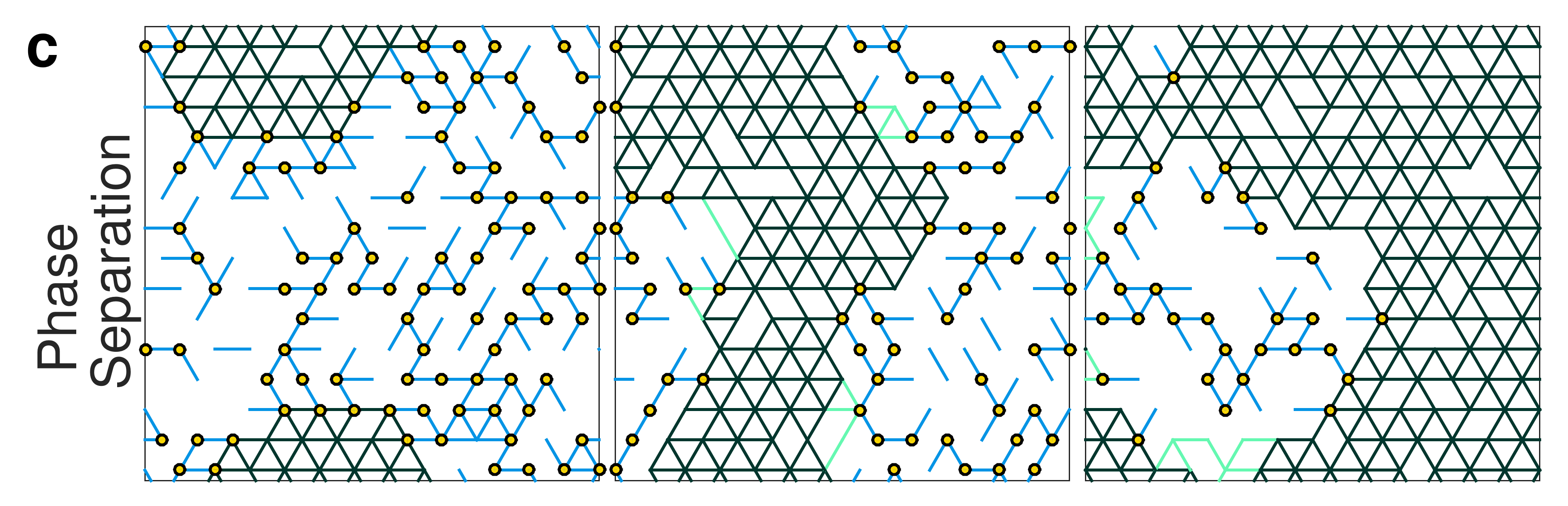}\\
\includegraphics[width=1.0\columnwidth]{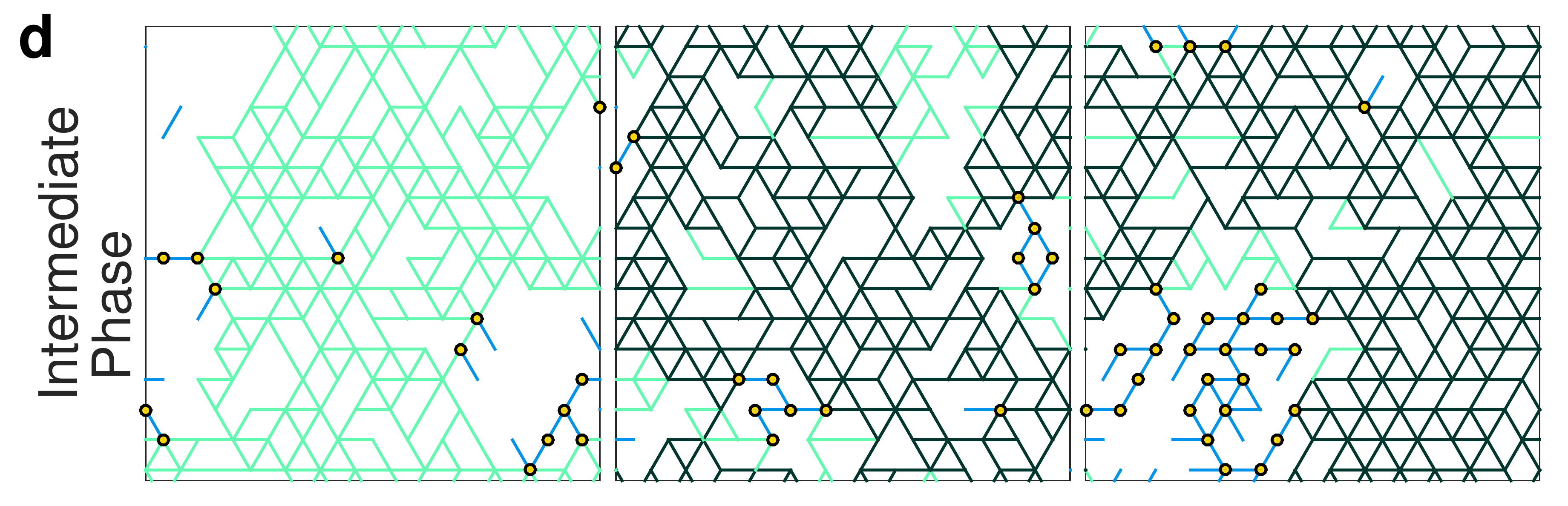}
\caption{\small{Typical network structures for (a) random networks, (b) self-organized networks, (c) entropy-favored networks, and (d) equilibrated networks at $T_g$. From left to right, are structures below, at, and above the rigidity transition. For illustration, we implement the pebble game algorithm~\cite{Jacobs95,Jacobs97} to decompose the networks into irreducible rigid clusters, unstressed (blue and green) and stressed (dark green), which are connected by pivots~\cite{Ellenbroek15}, shown as yellow circles.  
Green colors the connections in the percolating cluster, and the remainder clusters make the floppy regions in blue. 
}}\label{phase}
\end{figure}

Chalcogenides, for example, are network glasses composed of chemical elements with different covalent valences {$r$, proportional to which the number of covalent constraints $n$ varies}. 
Rather than a point threshold $r_c=2.4$~\cite{Phillips79,Thorpe85}, 
a range of singular features, named the intermediate phase, bridges the well-connected stressed and poorly coordinated floppy phases, as observed in experiments~\cite{Boolchand01,Rompicharla08,Bhosle12} and reproduced in molecular dynamics simulations~\cite{Micoulaut13,Bauchy14,Bauchy15}. 
Inside the phase, the non-reversible heat, a glass-transition equivalent of the latent heat, vanishes~\cite{Boolchand01}, which is associated with a vanishing stress heterogeneity~\cite{Rompicharla08} and a minimal molar volume~\cite{Bhosle12}. {All of these measurements are discontinuous } when entering the phase from either side~\cite{Bhosle12}. 
The critical point observed in random networks~\cite{Jacobs95,Jacobs96,Barre05}(Fig.~\ref{phase}(a)), which allow fluctuations in local connectivities, fails to capture the nature of the intermediate phase. 
Emerging in self-organized networks to reduce the energetic costs of self-stressed states~\cite{Thorpe00,Chubynsky06}(Fig.~\ref{phase}(b)), the rigidity window with distinct onsets of rigidity and self-stress promisingly maps to a critical range like the intermediate phase; however, the stronger heterogeneity inside the critical window actually contradicts the experimental observations, and the window is also sensitive to the appearance of prevailing perturbations such as van de Waals forces~\cite{Yan14}. 
In fact, {a rather odd feature} is the heterogeneous nature away from the threshold, outside of the intermediate phase. What causes the heterogeneity beyond the local fluctuations?

Recent achievements~\cite{Frenkel15,Escobedo14} indicate that the entropy, a synonym of ``disorder'', leads to order and heterogeneity in many cases, 
including the gas-crystal phase separation in colloid-polymer mixtures~\cite{Pusey86,Lekkerkerker92} and the open lattice structures of patchy particles~\cite{Smallenburg13,Mao13b}. 
The key components that allow for this comprehensive role are {\it (i)} the high degeneracy of configurations and {\it (ii)} the separation of degrees of freedom carrying entropy from the ones assembling structures.
In amorphous networks, configurations are inherently degenerate. Floppy and soft modes on boson peaks store significant amounts of vibrational entropy~\cite{Naumis05}, particularly close to {$n_c$}; thus, they inevitably shape the network structures. 
In this paper, we investigate the role of entropy in regulating network structures and show the appearance of phase-separated heterogeneous structures ruled by a critical point at the rigidity threshold. 
We then confirm the appearance of a homogeneous intermediate phase when stress energy dominates at low temperature. 
Finally, we present several experimental observations in chalcogenides. 

\section{Model}
\begin{figure}[h!]
\includegraphics[width=.55\columnwidth]{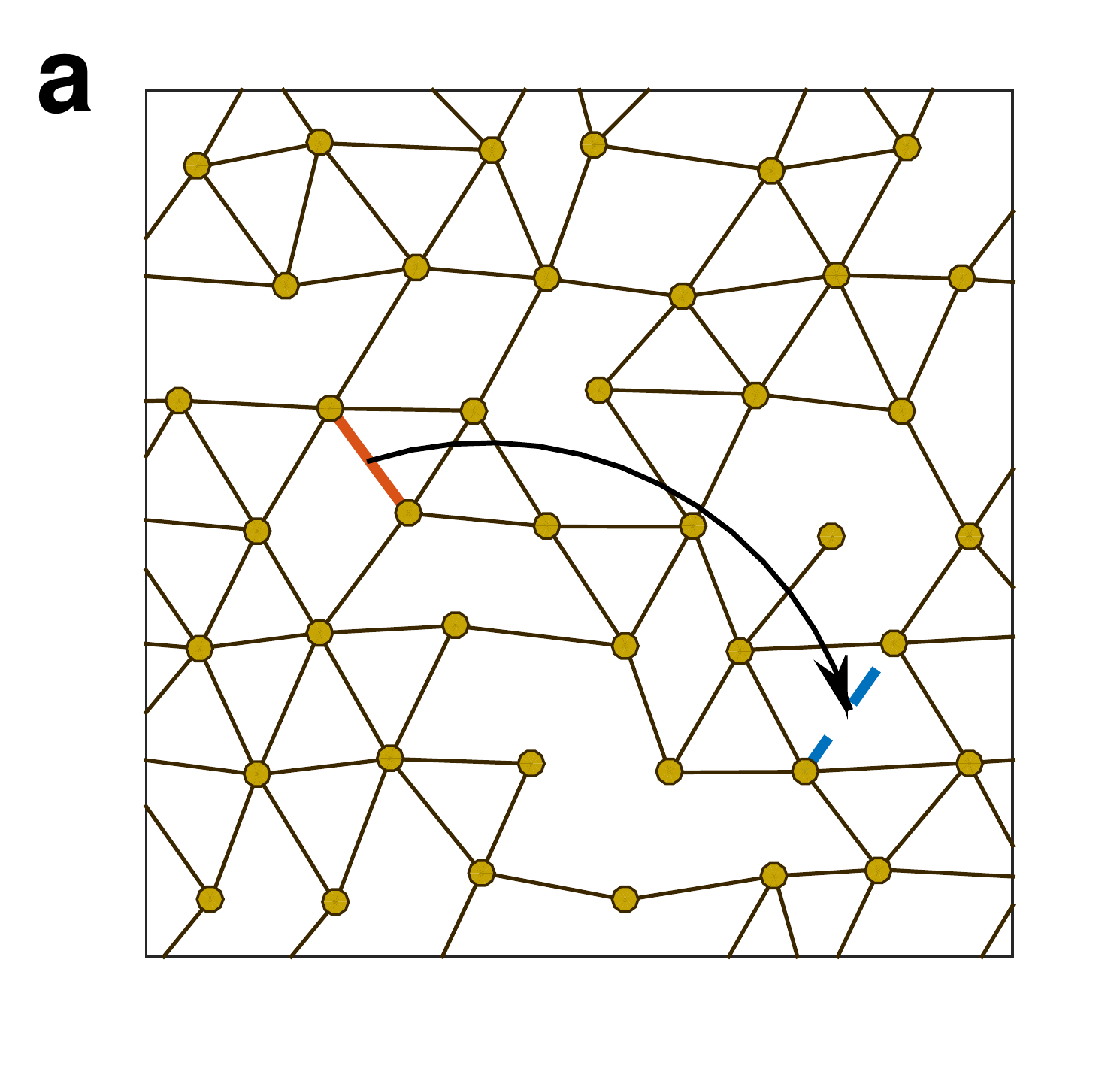}
\includegraphics[width=.43\columnwidth]{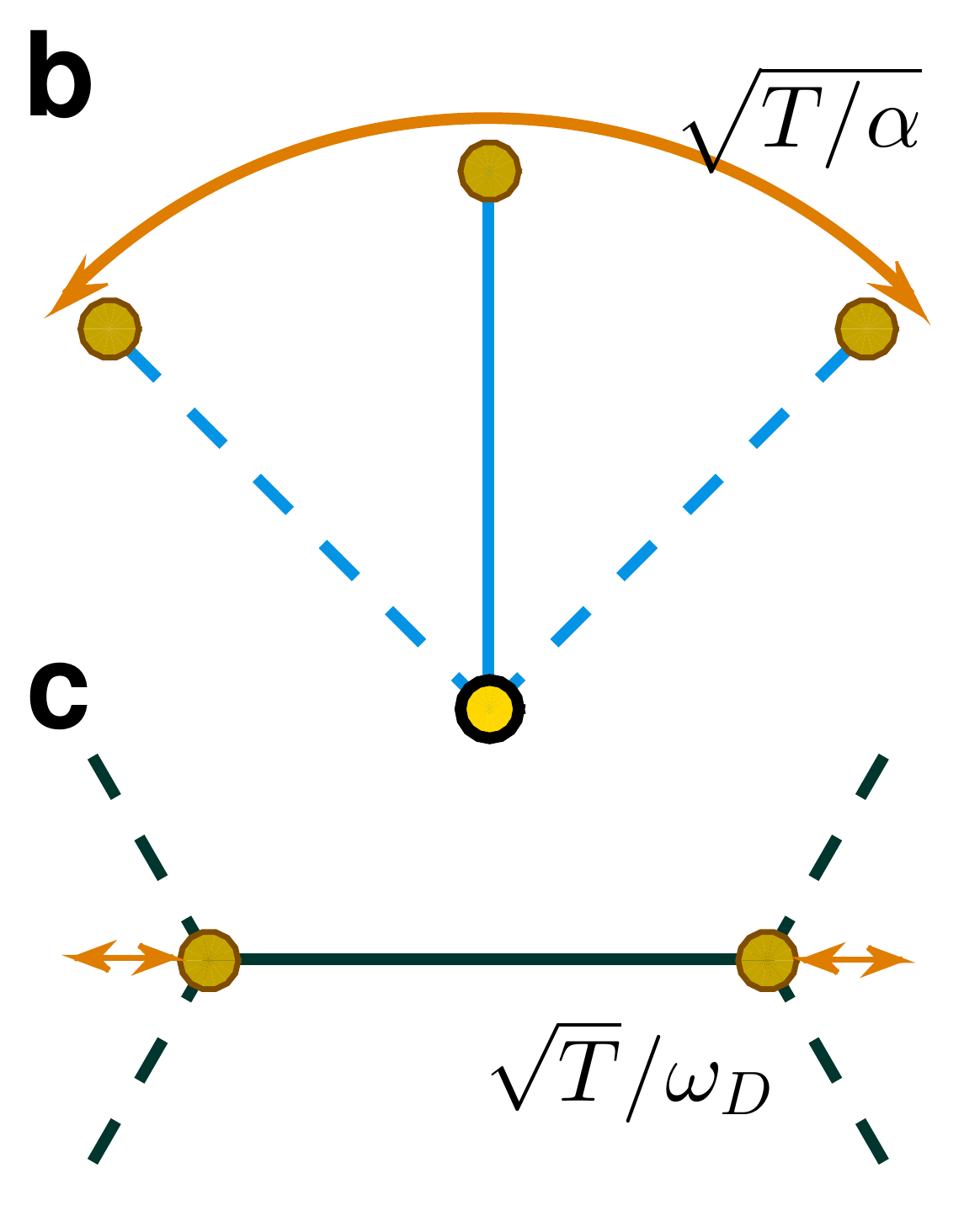}
\caption{\small{(a) Illustration of the network model on the triangular lattice. The configuration of spring connections between perturbed neighbor nodes, shown as solid lines, defines a network structure. Weak springs (not shown) connect all second neighbors. A new configuration is sampled by moving a randomly selected strong spring in red to a random vacant lattice link shown in blue dashed line. Illustrations of thermal vibrations corresponding to (b) a floppy mode and (c) a Debye-frequency mode.}}\label{model}
\end{figure}

We consider a network model 
on a two-dimensional triangular lattice with periodic boundaries. A particle at each of $N$ nodes can be wired to at most all of its six neighbors, {corresponding to the maximal constraint number $n_{m}=3$}. 
Following reference~\cite{Jacobs95}, we randomly perturb the locations of lattice nodes to avoid straight lines that lead to non-generic singular modes, as shown in Fig.~\ref{model}(a). 
{The key assumption of the model is the separation of energy scales such that we can consider the network of the stronger interactions such as the covalent bonds in chalcogenides and treat the weaker ones such as van der Waals forces as perturbations. In the simplest construction, a network configuration $\Gamma$ is defined by the allocation of {$N_{\rm s}\equiv nN$} linear springs of identical stiffness $k$ on the {$n_{m}N$} possible links.} 

Different configurations are probed by relocating one random spring (red solid) to an unoccupied (blue dashed) link at a time, as illustrated in Fig.~\ref{model}(a), such that their number is fixed by a given {average number of constraints $n$}, similar to rearranging atoms of different valences in network glasses. The different configurations are sampled with probabilities proportional to the Boltzmann factor $\exp(-\mf/T)$ using the Metropolis algorithm, which is documented together with the model parameters in {\it Methods}.  
Given configuration $\Gamma$, its free energy is 
\be
\mf(\Gamma) = \mh_0(\Gamma)-TS_{\rm vib}(\Gamma), 
\label{e4}
\ee 
where vibrational entropy $S_{\rm vib}$ quantifies the volume of thermal vibrations near the mechanical equilibrium of $\Gamma$~\cite{Naumis05,Mao13b,Mao15}, 
\be
\label{eq_svib}
S_{\rm vib}(\Gamma)=-\frac{1}{2}\ln\det\frac{\mm(\Gamma)}{T}\\=-\sum_{\omega}\ln\omega+c 
\ee 
which depends on $\omega^2$--the eigenvalues of Hessian $\mm$ and a $\Gamma$-independent number $c$. 
$\mh_0$ is the self-stress energy of $\Gamma$ at equilibrium.  
We introduce frustrations by imposing that 
the rest length of the spring $\gamma$ positioned at the link $\langle ij\rangle$, $l_\gamma=r_{\langle i,j\rangle}+\epsilon_\gamma$, differs from $r_{\langle i,j\rangle}$, the spacing between neighboring nodes $i$ and $j$, by a mismatch $\epsilon_\gamma$ assigned from a Gaussian distribution of zero mean and variance $\epsilon^2$. 
In the small frustration limit, where $\epsilon$ is much smaller than the lattice constant, we compute $\mm$ and $\mh_0$ in the linear approximation, as derived in the Supplementary Information Section A and Refs.~\cite{Yan13,Yan14,Yan15a}. 

{We include perturbations of non-specific but weaker interactions by connecting all {six} second neighbors on the lattice with springs of stiffness $k_{\rm w}\ll k$.} 
At this high connectivity, they act approximately as isotropic potentials of effective stiffness {$\alpha=\frac{6k_{\rm w}}{dk}\ll1$ time of $k$}. 
These weak forces hence set a finite vibration volume for floppy modes while leaving the other modes nearly untouched, as illustrated in Fig.~\ref{model}(b) and (c). 

\section{Results}
\begin{figure}[htbp]
\centering
\includegraphics[width=.96\columnwidth]{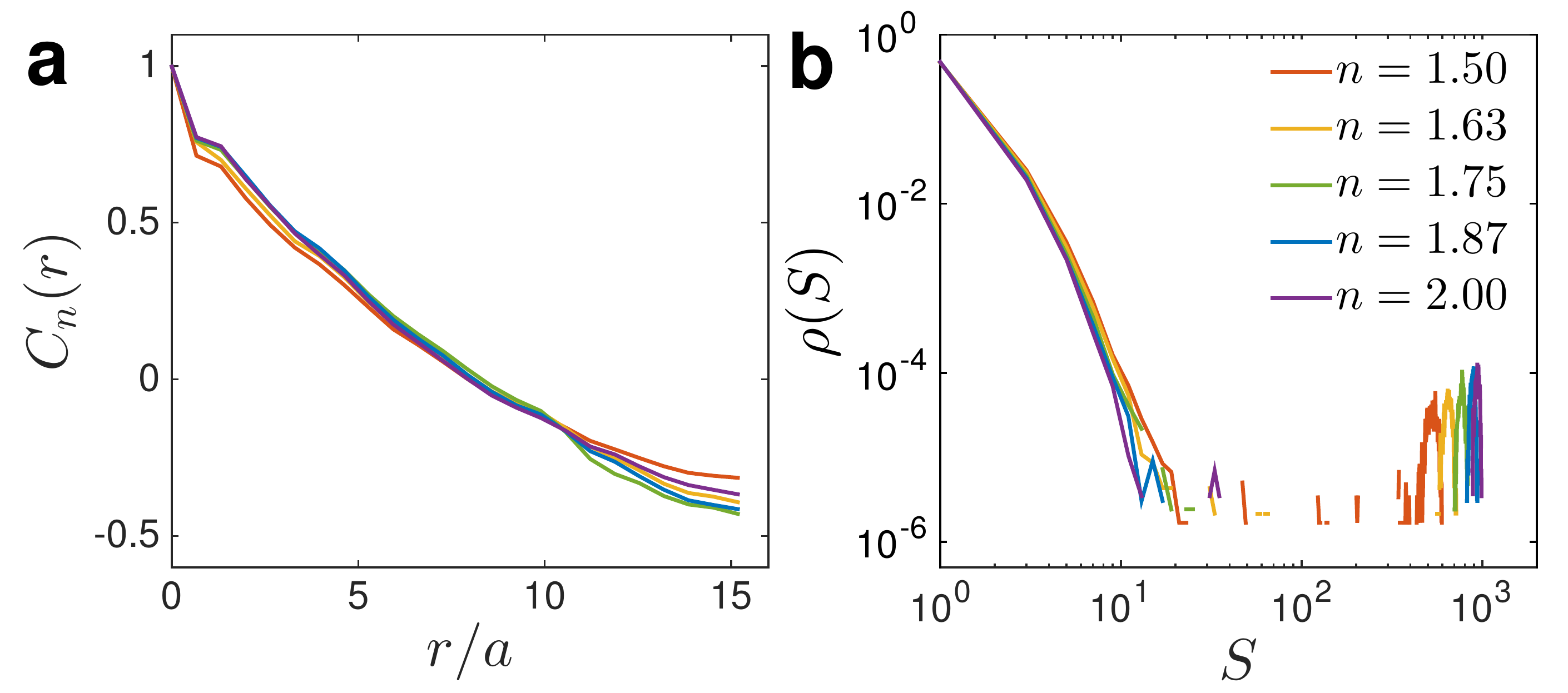}\\
\includegraphics[width=1.\columnwidth]{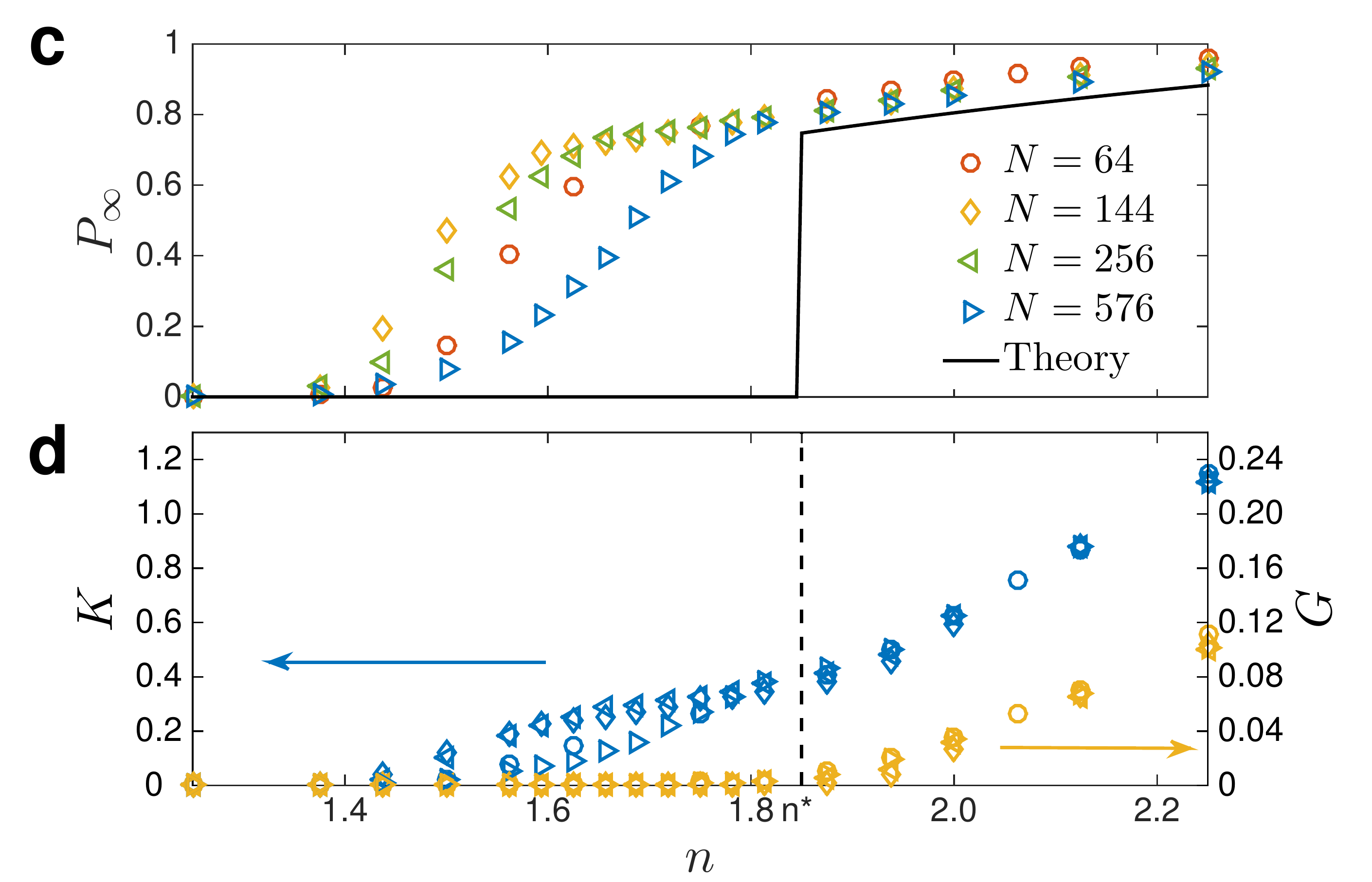}
\caption{\small{(a) Spatial correlation of {connectivity $C_n(r)=(\langle n(r)n(0)\rangle-\bar{n}^2)/(\langle n(0)^2\rangle-\bar{n}^2)$} for entropy favored networks, $N=576$. (b) Probability distribution of rigid cluster sizes $\rho(S)$,  {collapse for a wide range in $n<n_c$}.  (c) Probability in the percolating cluster $P_\infty$, and (d) Bulk modulus $K$ and shear modulus $G$ versus {constraint number $n$} for various system sizes $N$, $\alpha=0.0003$. The black solid line is theoretical prediction for the thermodynamic limit $N\to\infty$. {$\lambda\approx3.3$, so $n_f\approx0.94$, $n_r\approx2.76$, and $n^*\approx1.85$}, fitted by Eqs.(\ref{eq_sol}). 
}}\label{data}
\end{figure}

{\bf Entropy favors phase separation.} 
As shown in Fig.~\ref{phase}(c), in the limit of no self-stress penalty $\epsilon=0$ and thus no energy regulation $\mh_0=0$, 
entropy-favored networks present a phase separation into two phases, a highly coordinated stressed cluster ({$n>n_c$} red) and a floppy phase formed by the remainning clusters ({$n<n_c$} blue), near {$n_c$}, 
distinct from the homogeneous structures in Figs.~\ref{phase}(a) and (b), 
where the percolating rigid cluster would appear indistinguishable from the remainder if the color code and the pivots are removed in Fig.~\ref{phase}. 
This phase separation is captured by a long-range correlation of the local constraint number and a bimodal cluster size distribution (a system-size stressed cluster plus small ones in the floppy phase) in contrast to a continuous one~\cite{Souza09}, as shown in Fig.~\ref{data}(a) and (b). 

Due to the phase separation, the network rigidity arises in a discontinuous fashion as the stressed cluster percolates--growing from an island inside the floppy sea to a continent enclosing floppy lakes. 
This percolation occurs at a constraint number {$n^*$} different from {$n_c$}, 
which is captured by a discontinuous $P_{\infty}$, the probability of springs in the percolating cluster,  
as shown in Fig.~\ref{data}(c). In Fig.~\ref{data}(d), the bulk modulus $K$ shows a trend to jump at {$n^*$}, whereas the shear modulus $G$ vanishes. 

{\bf Phase diagram.} 
Why does entropy {alone} favor a floppy-rigid phase separation? 
As the degrees of freedom carrying vibrational entropy (particles) disconnect from the ones coding the configuration (springs), the total entropy increases by creating floppy modes in the floppy subpart of the network by confining springs in the stressed counterpart, 
particularly when this spring redistribution costs little configurational entropy near the rigidity threshold. 
{When the self-stress energy is not participating, the balance between the vibrational entropic gain and the configurational cost determines the stability of the separation.}

\begin{figure}[htbp]
\centering
\includegraphics[width=1.\columnwidth]{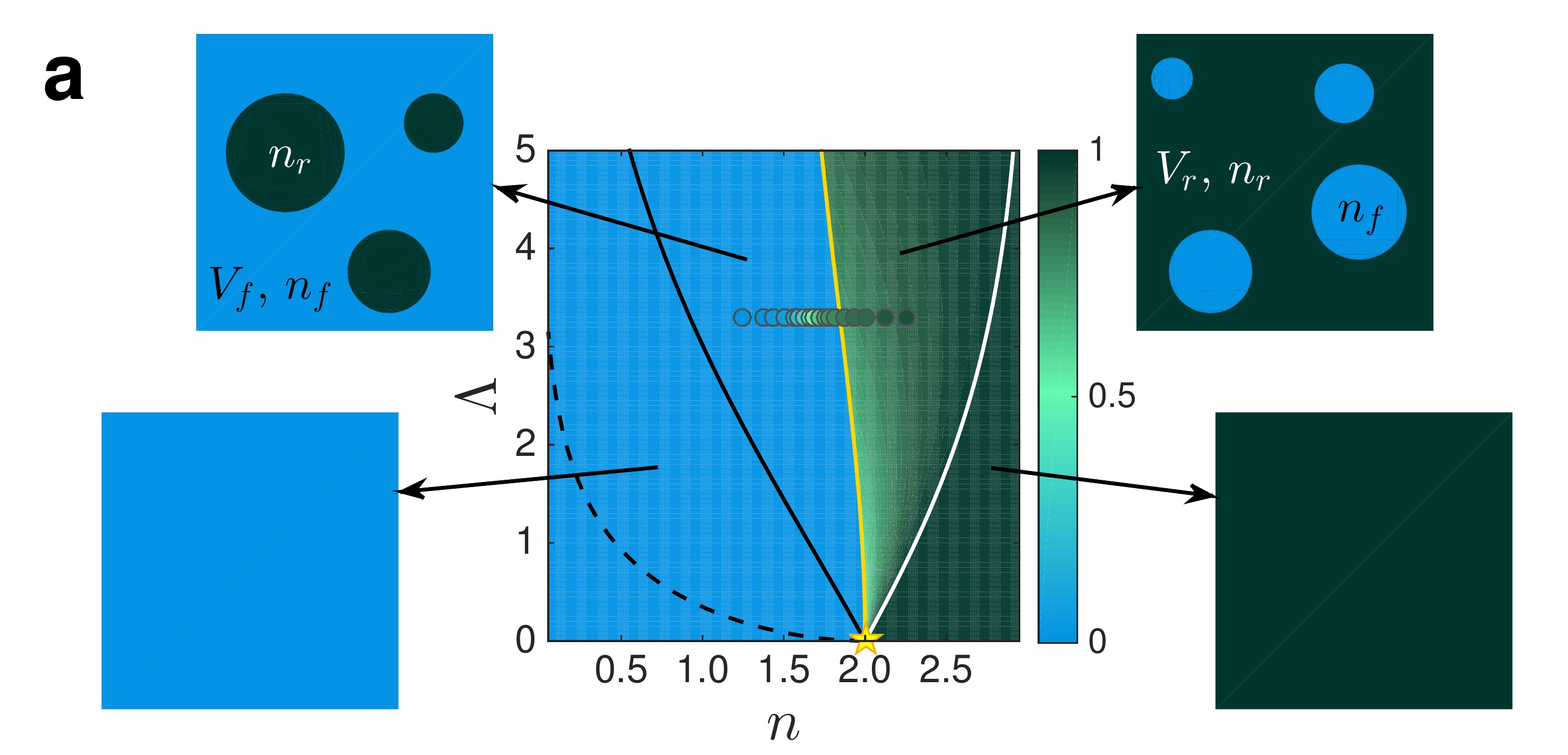}\\
\includegraphics[width=1.\columnwidth]{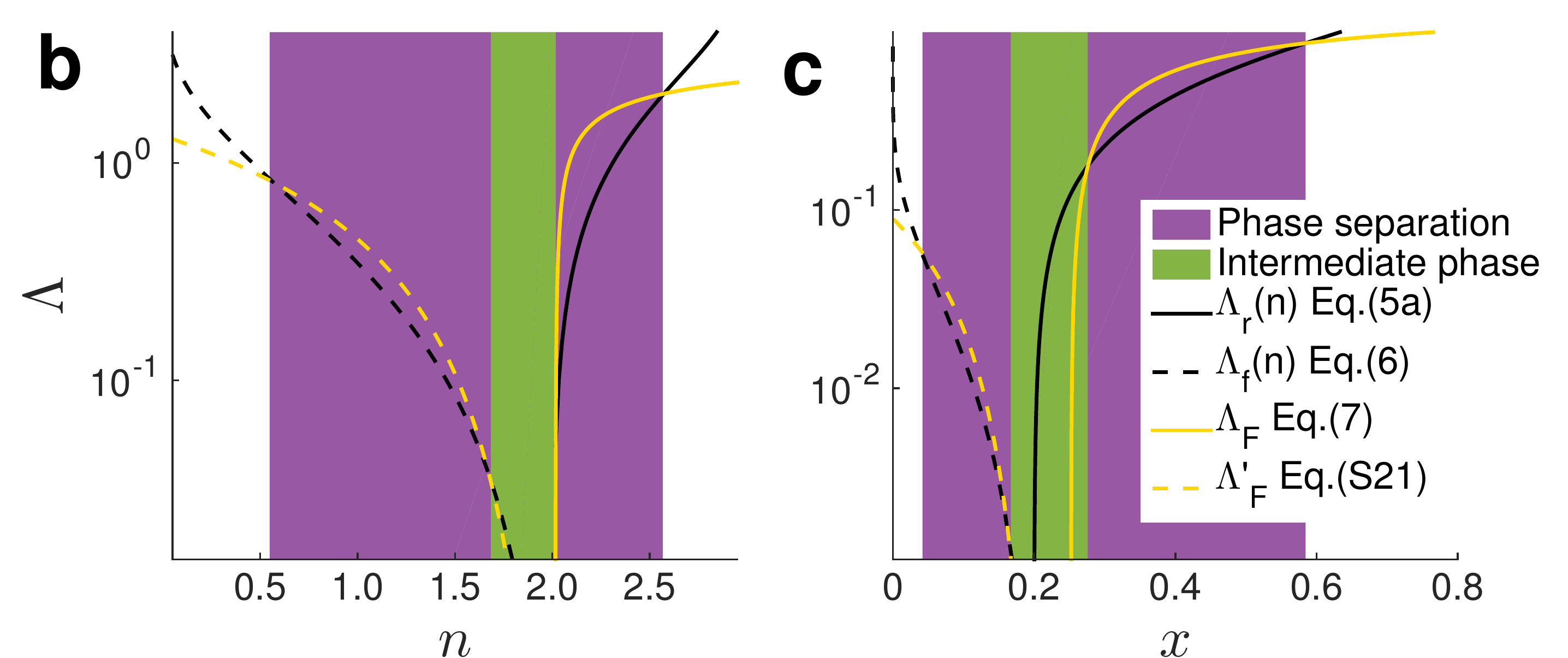}
\caption{\small{
(a) Phase diagram of the model in {$n$}-$\Lambda$ space. The star sign marks the critical point at {$n_c=d$}, $\Lambda_c=0$. The phase boundaries shown by black and white solid lines defined in Eqs.(\ref{eq_sol}) separate the heterogeneous phase mixed of floppy and stressed regions and homogeneous phases as illustrated. The floppy parts are in light blue and the stressed parts are in dark green. The dashed line shows the phase boundary Eq.(\ref{eq_RPzf}) towards a floppy-isostatic mixture phase. The color bar labels the probability of a bond in the percolating cluster $P_{\infty}$, which jumps at the yellow line {$n^*(\Lambda)$} when the rigid phase reaches half volume fraction. {Numerical data of $N=576$ $\lambda=3.3$ are shown in circles.}
(b) Phase diagram of model at $T_g$ with $\Lambda$ in log scale. The black solid and dashed line reproduce the phase boundaries (white solid and black dashed) in (a). {On the rigid side $n>n_c$, when the free energy loss at $T_g$ given by Eq.(\ref{eq_lb}) shown by yellow solid line is above the boundary, the heterogeneous networks appear in equilibrium. When $n<n_c$, the phase separation is stable when the yellow dashed line showing the free energy loss given by Eq.(S21) goes beyond the heterogeneous boundary.} 
(c) Same phase diagram showing the intermediate phase for compounds $A_xB_{1-x}$. {The purple regions show the range of heterogeneous phases, and the green region is the homogeneous intermediate phase.} 
}}\label{diag}
\end{figure}

Consider a separation into a homogeneous rigid phase and a floppy phase of volume fractions $V_r$ and $V_f$ controlled by the {constraint numbers $n_r$ and $n_f$, as illustrated in Fig.~\ref{diag}(a)}. 
The configurational entropy {is the entropy of mixing springs and vacancies summed over the two phases}, 
{
\begin{multline}
\frac{S_{\rm conf}}{N}=s_{c,0}+V_r\lp n_r\ln\frac{n_m}{n_r}+(n_m-n_r)\ln\frac{n_m}{n_m-n_r}\rp\\
+V_f\lp n_f\ln\frac{n_m}{n_f}+(n_m-n_f)\ln\frac{n_m}{n_m-n_f}\rp,
\label{eq_conf}
\end{multline}
plus $s_{c,0}$, the entropy from the boundary contribution, which vanishes in the thermodynamic limit.} 
{
As the extra vibrational entropy gains from the floppy modes, let us assume that the vibrational entropy is proportional to the number of floppy modes,}
{
\be
\frac{S_{\rm vib}}{N}=s_{v,0}+ V_f(n_c-n_f)\Lambda,
\label{eq_sv}
\ee} 
{changing by $\Lambda$ per floppy mode. As shown in the Supplementary Information Section B, this assumption is approximately valid in the model and per mode entropy gains $\lambda=-\frac{1}{2}\ln\alpha+\langle\ln\omega\rangle>0$,} where $\langle\ln\omega\rangle$ is the spectrum-average entropy of non-floppy modes. 
{Henceforce, we use the convention of the large $\Lambda$ as a parameter in the formalism and the small $\lambda$ as the actual entropic gain in the model.}

Constrained on the total volume $V_f+V_r=1$ and the average {constraint number $n_fV_f+n_rV_r=n$}, the total entropy $S_{\rm vib}+S_{\rm conf}$ is optimized with  
{
\begin{subequations}\label{eq_sol}
\begin{align}
&\frac{n_r}{n_m}=\frac{e^{-\frac{\Lambda n_c}{n_m}}-1}{e^{-{\Lambda}}-1};\label{eq_sola}\\
&\frac{n_f}{n_m}=\frac{1}{1+e^{\Lambda}(\frac{n_m}{n_r}-1)}=\frac{e^{\frac{\Lambda n_c}{n_m}}-1}{e^{{\Lambda}}-1};\\
&V_r=\frac{n-n_f}{n_r-n_f}.
\end{align}
\end{subequations}
}
Since $V_r\in[0,1]$, the heterogeneous phase exists in the self-consistent range {$n\in [n_f,n_r]$}, which is very wide {$\frac{n_r-n_f}{n_c}\sim\lambda\sim-\frac{1}{2}\ln\alpha$} for practical $\alpha$. 
The boundaries {$n_f(\Lambda)$} and {$n_r(\Lambda)$} 
define the heterogeneous separation phase in the phase diagram in Fig.~\ref{diag}(a). 

Analogous to the classical spontaneous magnetization and gas-liquid phase separation, the entropy-induced floppy-rigid separation is governed by a critical point at $\Lambda=0$ and {$n=n_c$}, 
but in a different universality class, as discussed in the Supplementary Information Section C. 
In the separation range, the network structure presents the dominant phase ($V>1/2$) with droplets of the subdominant one of a typical size characterized by the critical behavior approaching {$(n_c,0)$}. 
Global rigidity arises when the rigid phase becomes dominant at {$n^*=(n_f+n_r)/2$}, as indicated by the yellow line in Fig.~\ref{diag}(a). 

{\bf Self-stress prohibited.} 
When creating self-stressed states is prohibited~\cite{Thorpe00,Chubynsky06}, 
phase separation can still arise for {$n<n_c$} due to  
an entropy gain of additional soft modes on the boson peak in isostatic structures. 
Per degree of freedom in isostatic volume $V_c$, the vibrational entropy increases  
$
\Lambda'\equiv\frac{\partial S_{\rm vib}/N}{d\partial V_c}
$, 
positive as shown in the Supplementary Information Section B. 
This gain from isostatic structures leads to a separation between an isostatic phase and a floppy phase, as illustrated in Fig.~\ref{phase}J. 
The corresponding phase boundary follows 
{
\be
\Lambda'=\ln\frac{n_c}{n_f}+(\frac{n_m}{n_c}-1)\ln\frac{1-n_c/n_m}{1-n_f/n_m}
\label{eq_RPzf}
\ee
}
shown as the white dashed line in Fig.~\ref{diag}(a). 

{\bf Self-stress and homogeneous intermediate phase.} 
{Because reducing the self-stress energy tends to level the connection distribution~\cite{Yan14}, 
when the energetic cost $\mh_0$ competes with the entropic gain, 
a homogeneous intermediate phase can develop inside the heterogeneous gap at low temperature.} 
In Fig.~\ref{phase}(d), we depict the typical network structures equilibrating the total free energy Eq.(\ref{e4}) at the glass transition temperature $T_g$. From left to right, which correspond to below, at, and above $z_c$, the networks are floppy-isostatic heterogeneous, homogeneous, and floppy-stressed heterogeneous, respectively. 

{At temperature $T$} (in the energy unit $k\epsilon^2\equiv1$), each self-stressed state contributes an independent direction to store energy~\cite{Yan13,Yan15a}. 
Noticing the duality between self-stressed states and floppy modes~\cite{Kane14}, a free energy loss per floppy mode substitutes the entropy gain $\Lambda$ in Eq.(\ref{eq_sv}), 
\be
\Lambda\to\Lambda_F(T)=\Lambda-\frac{1}{2}\ln\lp1+\frac{1}{T}\rp
\label{eq_lb}
\ee
{(see the Supplementary Information Section D for the derivation)}. The self-consistent condition of floppy-rigid phase separation breaks down when 
{
$
\lambda_F(T)\leq\Lambda(n)
$}, 
the phase boundary in Eq.(\ref{eq_sola}). 
{Relying on the insights of the elastic models~\cite{Dyre06}, we apply a glass transition temperature that is proportional to the shear modulus, $T_g\propto G$, whose analytical form is derived in the Supplementary Information Section E.} {When $n>n_c$, $T_g\sim n-n_c$~\cite{Yan13,Yan15a}}, $\lambda_F$, shown as the blue solid line in Fig.~\ref{diag}(b), reenters the homogeneous phase when {$n$} decreases close to {$n_c$}, {$n_r-n_c\sim\alpha$}, defining the threshold of the homogeneous intermediate phase on the rigid side.

When {$n<n_c$}, $T_g\sim\alpha\ll1$~\cite{Yan13,Yan15a}, the self-stress prohibited situation applies. 
Derived from a flat mode density approximation~\cite{During13,Yan13} in the Supplementary Information Section D, the free energy loss per isostatic volume, shown as the blue dashed line in Fig.~\ref{diag}(b), 
surpasses the heterogeneous boundary Eq.(\ref{eq_RPzf}) in the dashed line {at $n_c-n_f\gtrsim\sqrt{\alpha}$}, 
giving the transition from the intermediate phase on the floppy side. 
Altogether, as the connectivity increases, the network structures change from homogeneous floppy to heterogeneous floppy-isostatic to intermediate homogeneous marginal to heterogeneous floppy-stressed and finally to homogeneous stressed, as depicted in Fig.~\ref{diag}(b). 

\section{Discussion}

{
{\bf Relative entropy.} 
This floppy-rigid phase separation has a general information theory implication. 
Rewriting the phase boundaries $n_f(\Lambda)$ and $n_r(\Lambda)$ in Eqs.(\ref{eq_sol}) in terms of relative entropies~\cite{Mezard09}, $D(p|q)=p\ln\frac{p}{q}+(1-p)\ln\frac{1-p}{1-q}$, we find that 
\begin{subequations}
\begin{align}
& D\lp\frac{n_c}{n_m}\right|\left.\frac{n_f}{n_m}\rp=D\lp\frac{n_c}{n_m}\right|\left.\frac{n_r}{n_m}\rp;\\
& (n_c-n_f)\Lambda  = n_m D\lp\frac{n_f}{n_m}\right|\left.\frac{n_r}{n_m}\rp.
\end{align}
\end{subequations}
The connection distributions of the floppy and rigid phases obey the conditions that (a) the relative entropy density from the rigid phase balances the density from the floppy one to the critical network 
and (b) the entropic gain per unit volume of the floppy phase compensates the relative entropy from the rigid phase to the floppy one.
Similarly, when any self-stress structure is forbidden, the phase boundary follows 
\be
n_c\Lambda=n_mD\lp\frac{n_c}{n_m}\right|\left.\frac{n_f}{n_m}\rp.
\ee
The entropic gain per unit volume of the critical structure compensates the relative entropy from the floppy phase to the critical phase. 

As derived and numerically verified in the Supplementary Information Section G, these balances, as well as the main results on the phase separation, hold in general for networks of multiple types of interactions, which is the case of real chalcogenides and proteins~\cite{Jacobs01}, as long as the vibrational entropy gain is approximately linear in probability distributions of interactions. 
}

{\bf Segregation in network glasses.} 
In network glasses, the degrees of freedom and the covalent constraints, both of which are associated with the atoms, depend differently on different chemical elements. The entropy-induced heterogeneous phase develops by segregating different elements. 
For illustration purposes, we derive in the Supplementary Information Section F the phase boundaries of compounds $A_xB_{1-x}$, where $x$ is the number fraction of atoms $A$, the knob equivalent to {the number of constraints $n$}. 
Particularly, we plot the phase diagram in Fig.~\ref{diag}(c) for chalcogenides $Ge_xSe_{1-x}$, where valences $r^{Se}=2$ and $r^{Ge}=4$ correspond to {the number of covalent constraints $n^{Se}=2$ and $n^{Ge}=7$ counting both bond-stretching and bond-bending contributions}~\cite{Thorpe85}. 
Segregations occur above the critical point ($\Lambda_c=0$ $x_c=0.2$), 
and five phases with four homogeneous-heterogeneous transitions appear at the glass transition in varying $x$. 

\begin{figure}[htbp]
\centering
\includegraphics[width=.7\columnwidth]{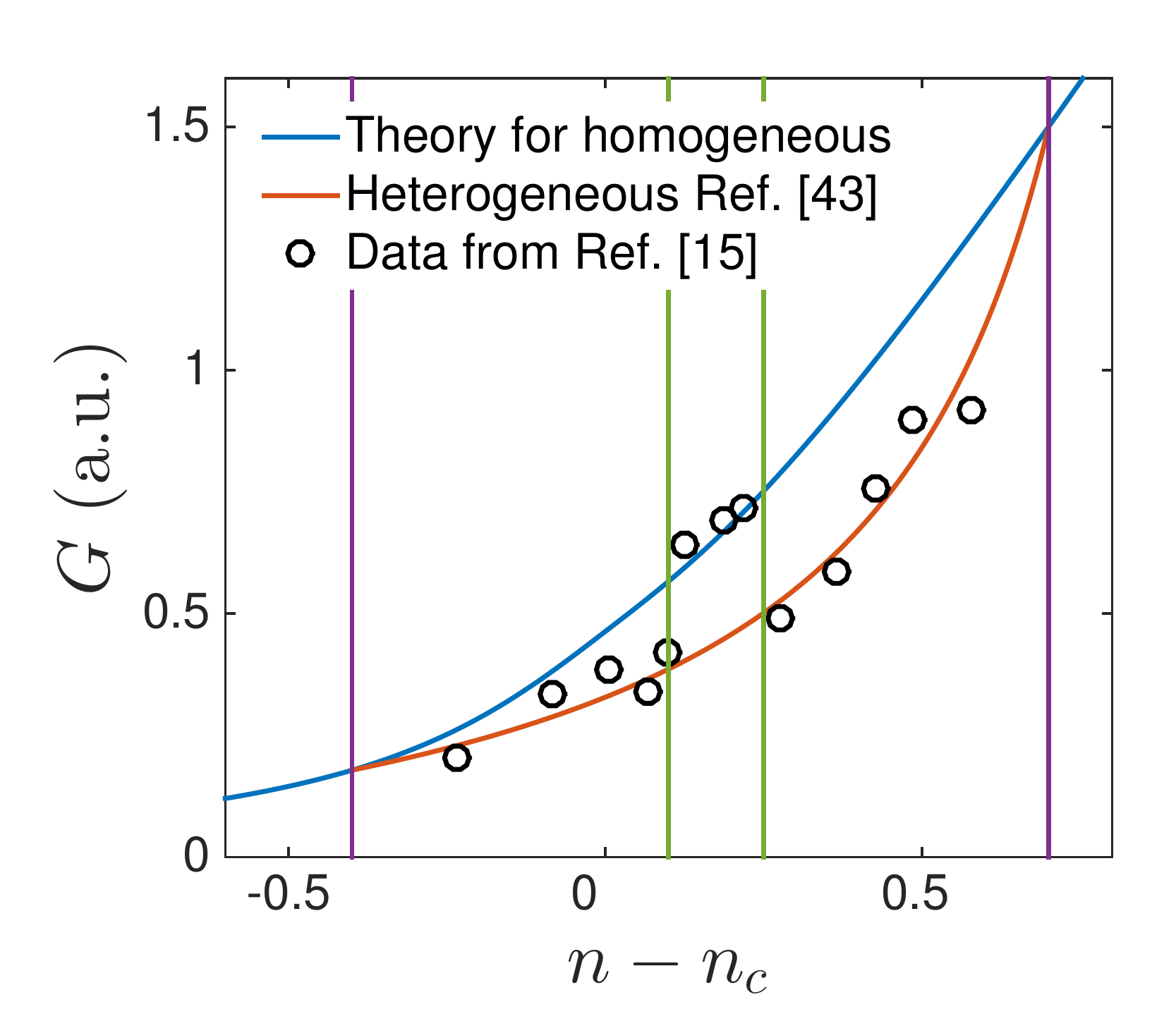}
\caption{\small{
Shear modulus $G$ in arbitrary unit predicted for homogeneous networks (blue) and for heterogenous networks composed of two phases at ends (red). Black circles are Raman shift data $\nu_{\rm TO}^2/\nu_0^2-1$ in Ref.~\cite{Rompicharla08}. The green and purple lines indicate the boundaries of intermediate phase and  separation range correspondingly, as in Fig.~\ref{diag}. 
}}\label{exp}
\end{figure}

{
{\bf Experimental indications on the intermediate phase and beyond.} 
This comprehensive structural behavior provides a natural interpretation for the four transitions with discontinuous features, including transitions to the intermediate phase, as observed in chalcogenides when changing the chemical compositions~\cite{Bhosle12}. 
Out of the intermediate phase, the micron-sized stress bubbles~\cite{Rompicharla08} are direct evidence of the heterogeneity. 
Its consequence on elasticity, the weakened shear modulus, is faithfully recorded in Raman scattering experiments~\cite{Rompicharla08}. 
Distortions of micro-structures shift the Raman peaks proportional to the global elasticity, $\Delta\nu^2\propto G$. %
As shown in Fig.~\ref{exp}, the jump of the Raman shift of the transversal optical branch in the intermediate phase~\cite{Rompicharla08} maps to the change of shear moduli between a homogeneous media and a heterogeneous mixture of two components~\cite{Hashin60}. 
In addition, high dynamical fragility out of the intermediate phase~\cite{Bhosle12} is consistent with the appearance of very floppy structures~\cite{Yan13}, and the Einstein relation breaks down with a floppy-phase-dominated diffusion and a stressed-phase-limited relaxation~\cite{Bauchy15}, which results in a very stretched exponential relaxation~\cite{Ediger96}. 

According to the model, ruling the transitions is predominantly the entropic gain $\lambda$, which is negatively correlated with $\alpha$, the strength of the perturbing interactions relative to that of the strong ones forming the network. The width of the heterogeneous range is $\Delta n\propto\lambda\sim-\frac{1}{2}\ln\alpha$, whereas that of the homogeneous intermediate phase is $\Delta n\sim\sqrt{\alpha}$. Thus the larger is the entropic gain, that is, in terms of experimental parameters, the stronger are the covalent bonds or the weaker are the van der Waals forces, the easier is the glass being frozen in a heterogeneous structure and the narrower is the intermediate phase. This rule provides a general reference to the component-dependent widths of the intermediate phase~\cite{Bauchy15}. 
Stabilizing the floppy parts as the weak interactions~\cite{DeGiuli14}, the pressure should be another experimentally approachable knob. Starting from a heterogeneous structure, increasing pressure effectively increases $\alpha$ and leads to a transition to the homogeneous phase~\cite{Bauchy14}. However, further pressure that distorts the strong interactions, $\alpha\sim1$, breaks our premise on the separation of energy scales and thus ends up in new physics~\cite{Bauchy15}. 
}



{\bf Conclusion.} 
We have shown that the entropy favors heterogeneous structures in the vicinity of the rigidity threshold of networks. 
Based on the counting approximation~\cite{Maxwell64,Kane14,Gohlke17}, we have derived a phase diagram for the network model and found that the critical point rules the phase separation. 
A homogeneous intermediate phase emerges inside the heterogeneous separation range when stress energy becomes dominant at low temperature. 
The resulting transitions among heterogeneous and homogeneous phases potentially resolve the discontinuous features of the intermediate phase in chalcogenides~\cite{Boolchand01,Rompicharla08,Bhosle12}. 
{The counting approximation simplifies the entropic gain as a single parameter independent of the configurations. To go further, it is necessary to treat the entropic gain more carefully and study the global minimum and the dynamics toward it in a rougher free energy landscape induced by the complex entropic consequences of structures such as long chains.}
Meanwhile, it is important to test the separation in molecular dynamics simulations~\cite{Micoulaut13} for various temperatures and non-specific weak forces. 
Finally, it is useful to apply the role of entropy in protein foldings and self-assembly, where flexible units appear vital for elastic functions~\cite{Yan17, Yan17a, Zheng17}. 


{\bf Methods.} 
We equilibrate network structures $\Gamma$ using the Metropolis algorithm. 
From an initial configuration $\Gamma$, a new configuration is proposed by the random relocation of a spring, as illustrated in Fig.~\ref{model}. By comparing the free energy Eq.(\ref{e4}) between the current and the new configurations, we sample and reset to the new configuration with probability $\min[1,\exp(-\frac{\mf(\Gamma')-\mf(\Gamma)}{T})]$, where parameter $T$ defines the equilibrated temperature. 
For each combination of parameters {$\{ n,T,\alpha\}$}, we implement in parallel 50 Monte Carlo simulations with $10^5$ steps to approach thermal equilibrium. 
When stress energy $\mh_0$ vanishes, $T$ is relevant only when thermal vibrations are so strong that Eq.(\ref{eq_sv}) breaks down and nonlinear terms become important, discussed in Supplementary Information Section H. 
In the model, we focus on the limit of the weak interactions $\alpha=0.0003$~\cite{Yan14,Yan15a}. 
In the segregation of chalcogenides, we apply $\alpha=0.03$, a choice closer to the actual strength of van der Waals forces~\cite{Yan13}. 
For the networks shown in Fig.~\ref{phase}(d), from left to right, they are equilibrated at {$n=1.625$}, $T=\alpha=0.0003$; {$n=2.0$}, $T=\alpha=0.0003$; and {$n=2.25$}, $T=0.1$. To illustrate the floppy-isostatic separation in the model, we amplify the free energy loss by six times, an artifact unnecessary for segregation in chalcogenides.  

\begin{acknowledgements}
I thank C.~Jian, J.~Liu, X.~Mao, B.~Shraiman and M.~Wyart for discussions, and anonymous referees for constructive suggestions. This work has been supported in part by the National Science Foundation under Grant No. NSF PHY 17-48958. I acknowledge support from the ``Center for Scientific Computing at UCSB" and NSF Grant CNS-0960316. 
\end{acknowledgements}

\bibliographystyle{unsrt}
\bibliography{Yanbib}

\begin{thebibliography}{10}

\bibitem{Hall03}
Randall~W. Hall and Peter~G. Wolynes.
\newblock Microscopic theory of network glasses.
\newblock {\em Phys. Rev. Lett.}, 90:085505, Feb 2003.

\bibitem{Shintani08}
Hiroshi Shintani and Hajime Tanaka.
\newblock Universal link between the boson peak and transverse phonons in
  glass.
\newblock {\em Nat Mater}, 7:870--877, Nov 2008.

\bibitem{Mauro09}
John~C Mauro, Yuanzheng Yue, Adam~J Ellison, Prabhat~K Gupta, and Douglas~C
  Allan.
\newblock Viscosity of glass-forming liquids.
\newblock {\em Proceedings of the National Academy of Sciences},
  106(47):19780--19784, 2009.

\bibitem{Yan13}
Le~Yan, Gustavo D{\"u}ring, and Matthieu Wyart.
\newblock Why glass elasticity affects the thermodynamics and fragility of
  supercooled liquids.
\newblock {\em Proceedings of the National Academy of Sciences},
  110(16):6307--6312, 2013.

\bibitem{Yan17}
Le~Yan, Riccardo Ravasio, Carolina Brito, and Matthieu Wyart.
\newblock Architecture and co-evolution of allosteric materials.
\newblock {\em PNAS}, 114:2526--2531, 2017.

\bibitem{Rocks17}
Jason~W Rocks, Nidhi Pashine, Irmgard Bischofberger, Carl~P Goodrich, Andrea~J
  Liu, and Sidney~R Nagel.
\newblock Designing allostery-inspired response in mechanical networks.
\newblock {\em Proceedings of the National Academy of Sciences},
  114(10):2520--2525, 2017.

\bibitem{Liu10}
Andrea~J. Liu, Sidney~R. Nagel, Wim van Saarloos, and Matthieu Wyart.
\newblock {\em The jamming scenario: an introduction and outlook}.
\newblock Oxford University Press, Oxford, 2010.

\bibitem{Maxwell64}
J.C. Maxwell.
\newblock On the calculation of the equilibrium and stiffness of frames.
\newblock {\em Philos. Mag.}, 27(5755):294--299, 1864.

\bibitem{OHern03}
Corey~S. O'Hern, Leonardo~E. Silbert, Andrea~J. Liu, and Sidney~R. Nagel.
\newblock Jamming at zero temperature and zero applied stress: The epitome of
  disorder.
\newblock {\em Phys. Rev. E}, 68(1):011306--011324, Jul 2003.

\bibitem{Silbert05}
L.~E. Silbert, A.~J. Liu, and S.~R. Nagel.
\newblock Vibrations and diverging length scales near the unjamming transition.
\newblock {\em Phys.\ Rev.\ Lett.}, 95:098301, 2005.

\bibitem{Wyart05b}
M.~Wyart.
\newblock On the rigidity of amorphous solids.
\newblock {\em Annales de Phys}, 30(3):1--113, 2005.

\bibitem{Phillips79}
J.C. Phillips.
\newblock Topology of covalent non-crystalline solids i: Short-range order in
  chalcogenide alloys.
\newblock {\em Journal of Non-Crystalline Solids}, 34(2):153 -- 181, 1979.

\bibitem{Thorpe85}
M.F. Thorpe.
\newblock Rigidity percolation in glassy structures.
\newblock {\em Journal of Non-Crystalline Solids}, 76(1):109 -- 116, 1985.

\bibitem{Boolchand01}
P~Boolchand, DG~Georgiev, and B~Goodman.
\newblock Discovery of the intermediate phase in chalcogenide glasses.
\newblock {\em Journal of Optoelectronics and Advanced Materials},
  3(3):703--720, 2001.

\bibitem{Rompicharla08}
K~Rompicharla, D~I Novita, P~Chen, P~Boolchand, M~Micoulaut, and W~Huff.
\newblock Abrupt boundaries of intermediate phases and space filling in oxide
  glasses.
\newblock {\em Journal of Physics: Condensed Matter}, 20(20):202101, 2008.

\bibitem{Bhosle12}
Siddhesh Bhosle, Kapila Gunasekera, Punit Boolchand, and Matthieu Micoulaut.
\newblock Melt homogenization and self-organization in chalcogenides-part ii.
\newblock {\em International Journal of Applied Glass Science}, 3(3):205--220,
  2012.

\bibitem{Micoulaut13}
Matthieu Micoulaut and Mathieu Bauchy.
\newblock Anomalies of the first sharp diffraction peak in network glasses:
  Evidence for correlations with dynamic and rigidity properties.
\newblock {\em physica status solidi (b)}, 250(5):976--982, 2013.

\bibitem{Bauchy14}
M~Bauchy, A~Kachmar, and M~Micoulaut.
\newblock Structural, dynamic, electronic, and vibrational properties of
  flexible, intermediate, and stressed rigid as-se glasses and liquids from
  first principles molecular dynamics.
\newblock {\em The Journal of chemical physics}, 141(19):194506, 2014.

\bibitem{Bauchy15}
M~Bauchy and M~Micoulaut.
\newblock Densified network glasses and liquids with thermodynamically
  reversible and structurally adaptive behaviour.
\newblock {\em Nature communications}, 6, 2015.

\bibitem{Jacobs95}
D.~J. Jacobs and M.~F. Thorpe.
\newblock Generic rigidity percolation: The pebble game.
\newblock {\em Phys. Rev. Lett.}, 75:4051--4054, Nov 1995.

\bibitem{Jacobs96}
D.~J. Jacobs and M.~F. Thorpe.
\newblock Generic rigidity percolation in two dimensions.
\newblock {\em Phys. Rev. E}, 53:3682--3693, Apr 1996.

\bibitem{Barre05}
J.~Barr\'e, A.~R. Bishop, T.~Lookman, and A.~Saxena.
\newblock Adaptability and ``intermediate phase'' in randomly connected
  networks.
\newblock {\em Phys. Rev. Lett.}, 94:208701, May 2005.

\bibitem{Thorpe00}
M.F Thorpe, D.J Jacobs, M.V Chubynsky, and J.C Phillips.
\newblock Self-organization in network glasses.
\newblock {\em Journal of Non-Crystalline Solids}, 266-269, Part 2(0):859 --
  866, 2000.

\bibitem{Chubynsky06}
M.~V. Chubynsky, M.-A. Bri\`ere, and Normand Mousseau.
\newblock Self-organization with equilibration: A model for the intermediate
  phase in rigidity percolation.
\newblock {\em Phys. Rev. E}, 74:016116, Jul 2006.

\bibitem{Yan14}
Le~Yan and Matthieu Wyart.
\newblock Evolution of covalent networks under cooling: Contrasting the
  rigidity window and jamming scenarios.
\newblock {\em Phys. Rev. Lett.}, 113:215504, Nov 2014.

\bibitem{Frenkel15}
Daan Frenkel.
\newblock Order through entropy.
\newblock {\em Nature materials}, 14(1):9--12, 2015.

\bibitem{Escobedo14}
Fernando~A Escobedo.
\newblock Engineering entropy in soft matter: the bad, the ugly and the good.
\newblock {\em Soft Matter}, 10(42):8388--8400, 2014.

\bibitem{Pusey86}
Peter~N Pusey and W~Van~Megen.
\newblock Phase behaviour of concentrated suspensions of nearly hard colloidal
  spheres.
\newblock {\em Nature}, 320(6060):340--342, 1986.

\bibitem{Lekkerkerker92}
HNW Lekkerkerker, WC-K Poon, PN~Pusey, A~Stroobants, and PB~Warren.
\newblock Phase behaviour of colloid+ polymer mixtures.
\newblock {\em EPL (Europhysics Letters)}, 20(6):559, 1992.

\bibitem{Smallenburg13}
Frank Smallenburg and Francesco Sciortino.
\newblock Liquids more stable than crystals in particles with limited valence
  and flexible bonds.
\newblock {\em Nature Physics}, 9(9):554--558, 2013.

\bibitem{Mao13b}
Xiaoming Mao, Qian Chen, and Steve Granick.
\newblock Entropy favours open colloidal lattices.
\newblock {\em Nature materials}, 12(3):217--222, 2013.

\bibitem{Naumis05}
Gerardo~G Naumis.
\newblock Energy landscape and rigidity.
\newblock {\em Physical Review E}, 71(2):026114, 2005.

\bibitem{Jacobs97}
Donald~J. Jacobs and Bruce Hendrickson.
\newblock An algorithm for two-dimensional rigidity percolation: The pebble
  game.
\newblock {\em Journal of Computational Physics}, 137(2):346 -- 365, 1997.

\bibitem{Ellenbroek15}
Wouter~G. Ellenbroek, Varda~F. Hagh, Avishek Kumar, M.~F. Thorpe, and Martin
  van Hecke.
\newblock Rigidity loss in disordered systems: Three scenarios.
\newblock {\em Phys. Rev. Lett.}, 114:135501, Apr 2015.

\bibitem{Mao15}
Xiaoming Mao, Anton Souslov, Carlos~I Mendoza, and Tom~C Lubensky.
\newblock Mechanical instability at finite temperature.
\newblock {\em Nature communications}, 6, 2015.

\bibitem{Yan15a}
Le~Yan and Matthieu Wyart.
\newblock Adaptive elastic networks as models of supercooled liquids.
\newblock {\em Physical Review E}, 92(2):022310, 2015.

\bibitem{Souza09}
Vanessa~K. de~Souza and Peter Harrowell.
\newblock Rigidity percolation and the spatial heterogeneity of soft modes in
  disordered materials.
\newblock {\em Proceedings of the National Academy of Sciences},
  106(36):15136--15141, 2009.

\bibitem{Kane14}
CL~Kane and TC~Lubensky.
\newblock Topological boundary modes in isostatic lattices.
\newblock {\em Nature Physics}, 10(1):39--45, 2014.

\bibitem{Dyre06}
Jeppe~C Dyre.
\newblock Colloquium: The glass transition and elastic models of glass-forming
  liquids.
\newblock {\em Reviews of modern physics}, 78(3):953--972, 2006.

\bibitem{During13}
Gustavo D{\"u}ring, Edan Lerner, and Matthieu Wyart.
\newblock Phonon gap and localization lengths in floppy materials.
\newblock {\em Soft Matter}, 9(1):146--154, 2013.

\bibitem{Mezard09}
Marc~and M\'ezard.
\newblock {\em Information, Physics and Computation}.
\newblock Oxford University press, 2009.

\bibitem{Jacobs01}
Donald~J Jacobs, Andrew~J Rader, Leslie~A Kuhn, and Michael~F Thorpe.
\newblock Protein flexibility predictions using graph theory.
\newblock {\em Proteins: Structure, Function, and Bioinformatics},
  44(2):150--165, 2001.

\bibitem{Hashin60}
Zvi Hashin.
\newblock {\em The elastic moduli of heterogeneous materials}.
\newblock US Department of Commerce, Office of Technical Services, 1960.

\bibitem{Ediger96}
M.~D. Ediger, C.~A. Angell, and S.~R. Nagel.
\newblock Supercooled liquids and glasses.
\newblock {\em J. Phys. Chem.}, 100:13200, 1996.

\bibitem{DeGiuli14}
Eric DeGiuli, Adrien Laversanne-Finot, Gustavo~Alberto D\"uring, Edan Lerner,
  and Matthieu Wyart.
\newblock Effects of coordination and pressure on sound attenuation, boson peak
  and elasticity in amorphous solids.
\newblock {\em Soft Matter}, 10(30):5628--5644, 2014.

\bibitem{Gohlke17}
Holger Gohlke, Ido~Y Ben-Shalom, Hannes Kopitz, Stefania Pfeiffer-Marek, and
  Karl-Heinz Baringhaus.
\newblock Rigidity theory-based approximation of vibrational entropy changes
  upon binding to biomolecules.
\newblock {\em Journal of Chemical Theory and Computation}, 2017.

\bibitem{Yan17a}
Le~Yan, Riccardo Ravasio, Carolina Brito, and Matthieu Wyart.
\newblock Principles for optimal cooperativity in allosteric materials.
\newblock {\em arXiv:1708.01820}, 2017.

\bibitem{Zheng17}
Yuanjian Zheng and Matthieu Wyart.
\newblock personal communication.

\bibitem{Calladine78}
C.R. Calladine.
\newblock Buckminster fuller's ``tensegrity'' structures and clerk maxwell's
  rules for the construction of stiff frames.
\newblock {\em International Journal of Solids and Structures}, 14(2):161 --
  172, 1978.

\bibitem{Wyart08}
M.~{Wyart}, H.~{Liang}, A.~{Kabla}, and L.~{Mahadevan}.
\newblock {Elasticity of Floppy and Stiff Random Networks}.
\newblock {\em Phys.\ Rev.\ Lett.}, 101:215501, 2008.

\bibitem{Derrida81}
Bernard Derrida.
\newblock Random-energy model: An exactly solvable model of disordered systems.
\newblock {\em Phys. Rev. B}, 24:2613--2626, Sep 1981.

\bibitem{DeGiuli14b}
Eric DeGiuli, Edan Lerner, Carolina Brito, and Matthieu Wyart.
\newblock Force distribution affects vibrational properties in hard-sphere
  glasses.
\newblock {\em Proceedings of the National Academy of Sciences},
  111(48):17054--17059, 2014.

\bibitem{Lindemann10}
FA~Lindemann.
\newblock {\em Z. Phys.}, 11:609, 1910.

\end{thebibliography}

\renewcommand{\theequation}{S\arabic{equation}}
\setcounter{equation}{0}
\setcounter{figure}{0}
\renewcommand{\thefigure}{S\arabic{figure}}

\begin{appendix}
\section{Supplementary Information}

\subsection{A. Linear approximation}
Consider  a  network of $N$ nodes connected by $N_c$ springs. 
If an infinitesimal displacement field $|\delta {\bf R}\rangle $ is imposed on the nodes, the change of length of the springs  can be written as a vector $|\delta r\rangle$ of dimension $N_c$. For small displacements, this relation is approximately linear: $|\delta r\rangle ={\cal S} |\delta {\bf R}\rangle $, where ${\cal S}$ is a $N_c\times Nd$ matrix.  To simplify the notation, we write ${\cal  S}$ as a $N_c\times N$ matrix of components of dimensions $d$, which gives ${\cal  S}_{\gamma,i}\equiv \partial r_\gamma/\partial {\bf R}_i=\delta_{\gamma,i} {\bf n}_\gamma$, where $\delta_{\gamma,i}$ is non-zero only if the spring $\gamma$ connects to the particle $i$, and  ${\bf n}_\gamma$ is the unit vector in the direction of the spring $\gamma$, pointing toward the node $i$.  Using the bra-ket notation, we can rewrite   ${\cal  S}=\sum_{\langle ij\rangle\equiv \gamma}| \gamma \rangle {\bf n}_{\gamma} (\langle i | -\langle j |)\nonumber$, where the sum  is over all the springs of the network.  Note that the transpose ${\cal S}^t$ of ${\cal  S}$ relates the set of contact forces $|f\rangle$  to the set $|{\bf F}\rangle$ of unbalanced forces on the nodes:  $| {\bf F}\rangle={\cal  S}^t |f\rangle$, which simply follows from the fact that ${\bf F}_i=\sum_{\gamma} \delta _{\gamma,i} f_\gamma {\bf n}_\gamma=\sum_{\gamma} f_\gamma {\cal S}_{\gamma,i}$ \cite{Calladine78}. 

{\bf Dynamic matrix.}
The dynamic matrix ${\cal M}$ is a linear operator connecting external forces to the displacements: ${\cal M}|\delta {\bf R}\rangle=|{\bf F}\rangle$. Introducing the $N_c\times N_c$ diagonal matrix ${\cal K}$, whose components are the spring stiffnesses ${\cal K}_{\gamma\gamma}=k_{\gamma}$, we have for harmonic springs $|f\rangle={\cal K} |\delta r\rangle$. Applying ${\cal  S}^t$ on each side of this equation, we get $|{\bf F}\rangle={\cal  S}^t |f\rangle={\cal  S}^t 
{\cal K} {\cal S} |\delta {\bf R}\rangle$, which thus implies \cite{Calladine78}:
\be
\label{4}
{\cal M}={\cal  S}^t {\cal K}{\cal  S}.
\ee
Note that in our model the diagonal matrix ${\cal K}$ contains two types of coefficients $k_{\rm w}$ and $k$, corresponding to the stiffnesses of weak springs and strong springs determining the configurations of networks. Then the  dynamic matrix can be written as ${\cal M}=k({\cal S}^t_{\rm s}{\cal S}_{\rm s}+\frac{k_{\rm w}}{k}\,{\cal S}^t_{\rm w}{\cal S}_{\rm w})$, where ${\cal S}^t_{\rm w}$ is the projection of the operator ${\cal S}^t$ on the subspace of weak springs. 
In the mean-field limit of weak interactions, {number of weak neighbors} $z_{\rm w}\rightarrow \infty$ while keeping $\alpha\equiv z_{\rm w} k_{\rm w}/(kd)$ constant, the weak springs lead to an effective interaction between each node and  the center of mass of the system~\cite{Wyart08}, so that, 
\be
{\cal M}\approx k\lp{\cal S}^t_{\rm s}{\cal S}_{\rm s}+\alpha{\cal I}\rp,
\label{alpha}
\ee
where ${\cal I}$ is a $dN\times dN$ identity matrix. 

Therefore, the vibrational modes of the strong network $|\delta{\bf R}_\omega\rangle$, 
\be
{\cal M}_{\rm s}|\delta{\bf R}_\omega\rangle=k\omega^2|\delta{\bf R}_\omega\rangle,
\label{omega}
\ee 
where ${\cal M}_{\rm s}={\cal S}_{\rm s}^t{\cal S}_{\rm s}$, are approximately the eigen vibrations of ${\cal M}$, 
\be
{\cal M}|\delta{\bf R}_\omega\rangle={\cal M}_{\rm s}|\delta{\bf R}_\omega\rangle+k\alpha |\delta{\bf R}_\omega\rangle=k(\omega^2+\alpha)|\delta{\bf R}_\omega\rangle
\ee
with the eigenvalues lifted up by $\alpha$. 

{\bf Stress energy.} 
The mismatches of link lengths to the spring rest lengths $|y\rangle$ generate an unbalanced force  field $|{\bf F}\rangle ={\cal  S}^t{\cal K}|y\rangle$ on the nodes, leading to a displacement $|\delta {\bf R}\rangle={\cal M}^{-1}{\cal  S}^t{\cal K}|y\rangle$. The elastic energy ${\cal H}=\frac{1}{2}\langle y-\delta r|\mathcal{K}|y-\delta r\rangle$ is minimal for this displacement and the corresponding energy ${\cal H}_0$ is:
\be
\label{5}
{\cal H}_0(|y\rangle)=\frac{1}{2}\langle y|{\cal K}-{\cal K}{\cal  S}{\cal M}^{-1}{\cal  S}^t{\cal K}| y\rangle.
\ee
In our model,  $y_\gamma=0$ for weak springs and $y_\gamma=\epsilon_\gamma$ is a Gaussian random variable for strong springs. Introducing ${\cal  S}^t_{\rm s}$, the operator ${\cal  S}^t$ on the subspace of strong springs of dimension $N_{\rm s}$, we have $k{\cal S}^t_{\rm s} |\epsilon\rangle\equiv {\cal  S}^t{\cal K}|y\rangle$ and Eq.(\ref{5}) becomes
\be
\label{6}
{\cal H}_0(|\epsilon\rangle)=\frac{k}{2}\langle\epsilon|{\cal I}-k{\cal S}_{\rm s}{\cal M}^{-1}{\cal S}^t_{\rm s}|\epsilon\rangle.
\ee

\subsection{B. Entropy gain $\lambda$}
In two phase separation, the vibrational entropy reads
{
\begin{multline}
\frac{S_{\rm vib}}{N}=-V_f(n_c-n_f)\ln\omega_0-n_fV_f\int\rd\omega D_f(\omega)\ln\omega\\-dV_r\int\rd\omega D_r(\omega)\ln\omega. 
\label{eq_vib}
\end{multline}
}
$\omega_0$ is the vibrational frequency of floppy modes, $\omega_0=\sqrt\alpha>0$ thanks to the weak interactions. 
In a weak field $\alpha\ll1$, the other mode frequency is leveled up approximately as $\omega'=\sqrt{\omega^2+\alpha}$. Density of non-floppy modes $D(\omega)\equiv\sum_{\omega'>\omega_0}\delta(\omega'-\omega)/\sum_{\omega'>\omega_0}$, $D_f(\omega)$ and $D_r(\omega)$ are densities of floppy and rigid phases accordingly. 

\begin{figure}[htbp]
\centering
\includegraphics[width=1.0\columnwidth]{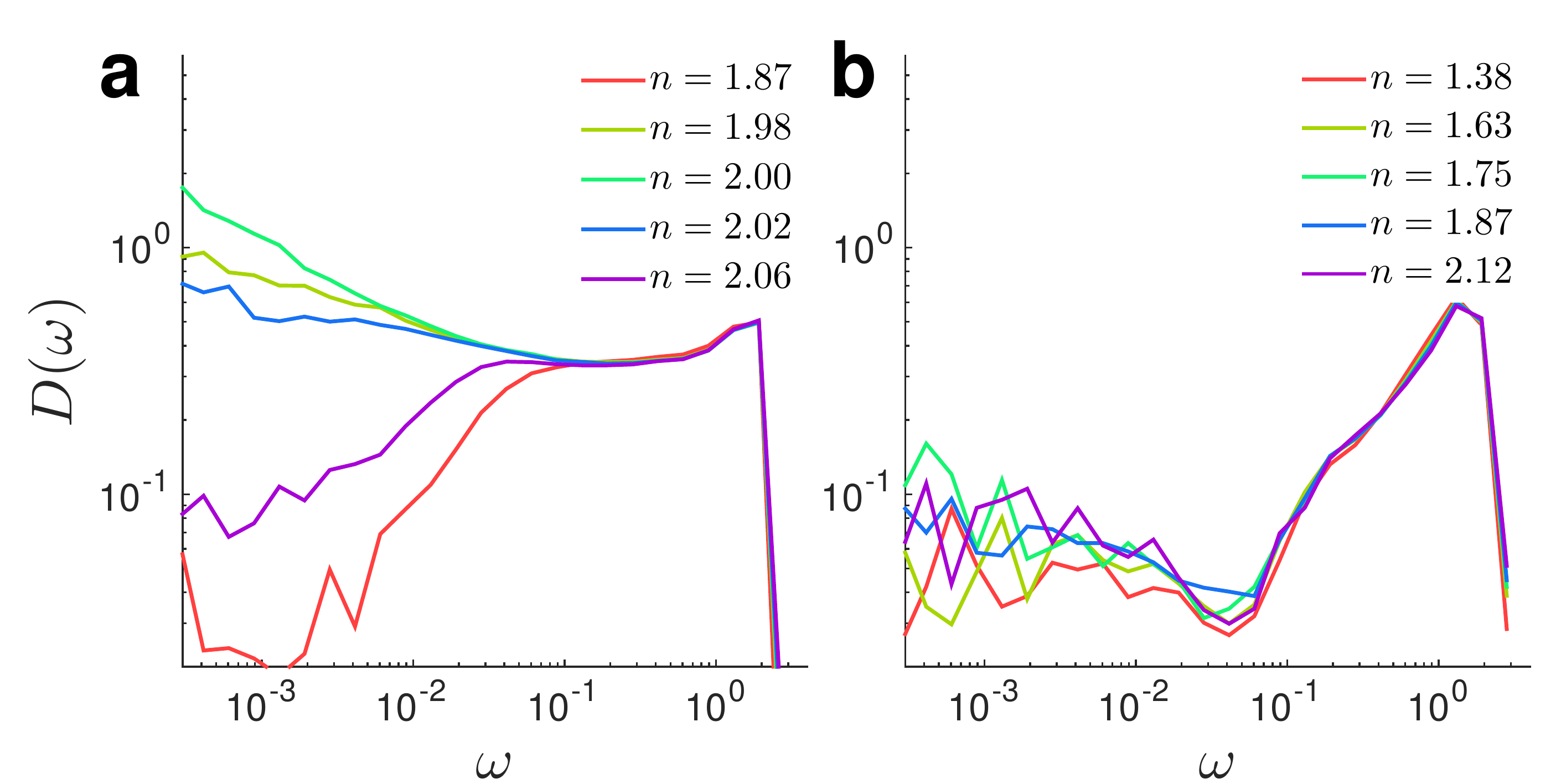}\\
\caption{\small{Density of vibrations $D(\omega)$ for (a) self-organized homogeneous networks and (b) phase separated networks $\alpha=0.0003$ with various number of constraints below and above the rigidity transition, $N=576$. Vibrational frequencies are computed without considering the weak forces.}}\label{dos}
\end{figure}

The densities of states of entropy-favored networks are shown together with ones of homogeneous networks in Fig.~\ref{dos}. 
As predicted by the mean-field theory, the density of homogeneous networks is cutoff on the low frequency end at the boson peak {$\omega^*\sim|n-n_c|$}, which is singular at the rigidity transition. 
For heterogeneous networks, the density of states presents no such singularity at the transition {$n^*$}. Like very stressed networks and very floppy ones, the density is blocked in high frequency modes cutoff at $\omega\sim0.1$. An odd feature is the appearance of low frequency modes, shown as a flat distribution with quite low density. We speculate the feature is related to a tendency of small clusters organizing into one dimensional chains. 

Though we have seen a significant change in the volume portion of the rigid phase and the floppy phase in the range of constraints number we prob, the densities of states for different numbers of constraints lay over on each other quite well, which implies that our approximation in Eq.(4) neglecting the difference between $D_f$ and $D_r$ is a good approximation.

{\bf Positive definiteness of $\lambda$.} 
Consider creating a floppy mode in the floppy phase, 
\be
{\Delta S_{\rm vib}} = -\ln\omega_0-Nd\int_{\omega_0}^{\omega_D}\rd\omega \Delta D(\omega)\ln\omega,
\label{eq_ds}
\ee
where the first term is the contribution from the floppy mode, while $\Delta D(\omega)$ in the second term includes the density shift of both $D_f$ and $D_r$ towards the Debye frequency $\omega_D$ when lowering the connectivity in the floppy phase and increasing {$n$} in the rigid phase~\cite{Wyart08,During13}. By definition, the total number of modes does not change, $Nd\int\rd\omega \Delta D(\omega)=-1$. We decompose the variance of density of states in a special way
\[
Nd\Delta D(\omega)=\rho_+(\omega)-\rho_-(\omega), 
\]
that both $\rho_+(\omega)$ and $\rho_-(\omega)\geq0$ for $\forall\omega\in[\omega_0,\omega_D]$, and $\int\rd\omega\rho_-(\omega)\omega=\omega_0+a\omega_D$, where $a\equiv\int\rd\omega\rho_+(\omega)$, so $\int\rd\omega\rho_-(\omega)=1+a$.
Then 
\begin{subequations}
\begin{align}
\Delta S_{\rm vib}&=-\ln\omega_0-\int\rd\omega\rho_+(\omega)\ln\omega+\int\rd\omega\rho_-(\omega)\ln(\omega)\\
&\geq-\ln\omega_0-a\ln\omega_D+\int_{\omega_0}^{\omega_D}\rd\omega\rho_-(\omega)\ln\omega\\
&\geq-\ln\omega_0-a\ln\omega_D+(1+a)\ln\omega_0\\
&+\frac{\ln\omega_D-\ln\omega_0}{\omega_D-\omega_0}\lp\int\rd\omega\rho_-(\omega)\omega-(1+a)\omega_0\rp\\
&=0.
\end{align}
\end{subequations}
In the second inequality, we have used the concaveness of $\ln\omega$, where the integral of $\ln\omega$ is larger than the integral of a linear function connecting the two end points.
Defined as vibrational entropy gain per floppy mode, 
{
\be
\lambda\equiv\frac{\partial S_{\rm vib}}{\partial N_f}\geq0.
\label{eq_lamb}
\ee
}

{\bf Self-stress prohibited.} The entropy increases by creating isostatic region {$\lambda'\geq0$}. By definition,
{
\begin{multline}
dN\lambda'=-\int_0^\infty\rd\omega [{\cal N}_c(\omega)-{\cal N}_f(\omega)]\ln\omega\\+(n_c-n_f)\int_0^\infty\rd\omega\partial_{n_f}{\cal N}_f(\omega)\ln\omega\\
=-\int_0^\infty\rd\omega\int_{n_f}^{n_c}\rd n\int_{n_f}^{n}\rd n'\partial_{n'}^2{\cal N}_{n'}(\omega)\ln\omega
\end{multline}
}
where {${\cal N}_n(\omega)=nN D(\omega)$} counts the number of vibrations $\omega$ for homogeneous network of {constraint number $n$}. As {$n$} increases by {$\rd n>0$ for $n<n_c$}, about {${N\rd n}$} vibrations emerge at {$\omega\sim n_c-n$}~\cite{During13}, {${\cal N}_{n+\rd n}(\omega)-{\cal N}_n(\omega)\geq0$} for $\forall \omega$. Equivalently, {$\partial_n{\cal N}_n(\omega)\approx b\partial_{\omega}{\cal N}_n(\omega)$}, with $b>0$. Therefore,
{
\be
dN\lambda'\approx-b^2\int_0^\infty\rd\omega\int_{n_f}^{n_c}\rd n\int_{n_f}^{n}\rd n'{\cal N}_{n'}(\omega)\partial_\omega^2\ln\omega\geq0
\ee
}
The inequality is again the result of concaveness of log function, $\partial_\omega^2\ln\omega<0$.

Specifically, we consider an approximation to the density of states in random networks of {constraint number $n$}: a flat density $D_0(\omega)=1/b$ cut off at {$\omega^*=b(n_c-n)$ and $\omega_D=bn_c$}~\cite{During13}, where $b\sim1$. 
{
\be
\frac{S_{\rm vib}}{Nd}\approx -\frac{V_c}{2\omega_D}\int_0^{\omega_D}\rd\omega\ln(\omega^2+\alpha)-\frac{V_f}{2\omega_D}\int_{b(n_c-n_f)}^{\omega_D}\rd\omega\ln(\omega^2+\alpha).
\label{eq_RPs0}
\ee
\begin{multline}
\lambda'%
=\frac{1}{2\omega_D}\int_0^{b(n_c-n_f)}\rd\omega\ln\frac{b^2(n_c-n_f)^2+\alpha}{\omega^2+\alpha}\\
=\frac{1}{\omega_D}\left[b(n_c-n_f)-\sqrt{\alpha}\arctan\frac{b(n_c-n_f)}{\sqrt{\alpha}}\right]\geq0.
\label{eq_RPlb}
\end{multline}
}
{$\lambda'\approx\frac{b^3}{3\alpha\omega_D}(n_c-n_f)^3$ for $n_c-n_f\lesssim\sqrt{\alpha}/b$ and $\lambda'\approx\frac{b}{\omega_D}(n_c-n_f)$ for $n_c-n_f\gtrsim\sqrt{\alpha}/b$}. 

\subsection{C. Universality class of floppy-rigid separation}
In our system, the entropy gain $\Lambda$ plays as the relevant parameter, like temperature $T-T_c$, while the average {constraint number $n$} as order parameter, similar to mean magnetization $M$ in ferromagnetic transition or mean density $\rho$ in gas-liquid separation. 
We can thus study the universality class of floppy-rigid separation by defining critical exponents mapping to the standard Landau theory of critical phenomena. 
Close to the critical point $\Lambda=0$ and {$n=n_c$}, the free energy follows, 
\be
\mf=-TS_{\rm vib}\sim\Lambda^{2+\alpha}.
\label{eq_falpha}
\ee
Inserting the counting approximation Eq.(4), we find $\alpha=-1$. 
The order parameter scales as, 
{
\be
n_{r,f}\sim\Lambda^{\beta}.
\label{eq_zbeta}
\ee }
The mean-field solution Eq.(5) implies $\beta=1$. Both exponents are different from the standard Landau theory. 

{

\subsection{D. Free energy at $T$}
For simplicity, we consider the annealed free energy $F_{\rm ann}=-T\ln\overline{\mz}$. It is exact in the random energy model~\cite{Derrida81} above the ideal glass transition~\cite{Mezard09} and we find it to be a good approximation of $F$ in network models~\cite{Yan13}. 
The over-line implies an average over quenched disorder $\epsilon$, 
\be
\overline{\mz}=\sum_{\Gamma}\overline{\exp[-\mf(\Gamma)/T]}^{\epsilon}.
\label{partfunc}
\ee
Applying the linear approximation Eq.(\ref{6}) and the Gaussian distribution $\rho(\epsilon_{\gamma})=\frac{1}{\sqrt{2\pi\epsilon^2}}e^{-\epsilon_{\gamma}^2/2\epsilon^2}$ of frustration at bond $\gamma$, we have 
\be
\overline{\mz}=\sum_{\Gamma}\exp\left[-\frac{1}{2}\tr\ln\left(\mi+\frac{\mg(\Gamma)}{T}\right)+S_{\rm vib}(\Gamma)\right].
\label{partz}
\ee
where we have used $k\epsilon^2=1$ to scale the temperature. As shown in Eq.(\ref{g0}), when $\alpha=0$, coupling matrix $\mg$ acts as a projection operator onto the null space of structure matrix $\ms_{\rm s}$. So
\be
-\frac{F}{NT}=\frac{S_{\rm conf}}{N}+\frac{S_{\rm vib}}{N}-\frac{V_r(n_r-n_c)}{2}\ln\lp1+\frac{1}{T}\rp,
\ee
for each self-stress direction created, free energy decreases by $\lambda_F=\lambda-\frac{1}{2}\ln(1+\frac{1}{T})$.

Including the perturbation $\alpha>0$ in the floppy region $n<n_c$, the total free energy for isostatic-floppy separation~\cite{Yan13} then follows 
{
\begin{multline}
\frac{F}{NT}=\frac{V_cn_c}{2}\int\rd\omega D_c(\omega)\ln\lp1+\frac{\alpha}{\alpha+\omega^2}\frac{1}{T}\rp\\
+\frac{V_fn_f}{2}\int\rd\omega D_f(\omega)\ln\lp1+\frac{\alpha}{\alpha+\omega^2}\frac{1}{T}\rp\\
+\frac{n_c-n}{2}\ln\lp1+\frac{1}{T}\rp-\frac{S_{\rm vib}}{N}-\frac{S_{\rm conf}}{N}.
\label{eq_RPf}
\end{multline}
}
So the free energy loss,
{
\begin{multline}
\lambda'_F=-\frac{\partial F/NT}{d\partial V_c}\approx\lambda'-\frac{1}{2\omega_D}\int_0^{b(n_c-n_f)}\rd\omega\ln\frac{1+\frac{\alpha/T}{\alpha+\omega^2}}{1+\frac{\alpha/T}{\alpha+b^2(n_c-n_f)^2}}\\
=\frac{1}{\omega_D}\lp b(n_c-n_f)-\sqrt{\alpha(1+\frac{1}{T})}\arctan\frac{b(n_c-n_f)}{\sqrt{\alpha(1+1/T)}}\rp,
\label{eq_fl}
\end{multline}
}
becomes approximately linear in {$n_c-n_f$ when $n_c-n_f\gtrsim\sqrt{\alpha}/b$}, 
faster than the heterogeneous boundary {$\Lambda\sim(n_c-n_f)^2$} in Eq.(6). 

\subsection{E. Shear modulus of perturbed networks}
We consider elastic model approximation $T_g\sim G$~\cite{Dyre06} for the glass transition temperature $T_g$. Here, we derive the scaling relations of $G$, $n$ and $\alpha$ from a perturbation theory. 
In the linear approximation Eq.(\ref{5}), the elastic energy $\mh_0$ is quadratic to any associated deformation $|y\rangle$. For a shear in $x$-$y$ plane, 
\[
|y\rangle=\gamma\left|\frac{\Delta x\Delta y}{\Delta r}\right\rangle,
\]
where $\gamma$ is the shear strain, $\Delta x$, $\Delta y$ and $\Delta r$ are the projection onto $x$ and $y$ directions and the length of the corresponding springs. 

Shear modulus of a configuration $\Gamma$,
\be
\label{gg}
G(\Gamma)=\frac{1}{V}\frac{\partial^2\mh_0(\Gamma)}{\partial \gamma^2}=\frac{1}{V}\left\langle\frac{\Delta x\Delta y}{\Delta r}\right|\mg\left|\frac{\Delta x\Delta y}{\Delta r}\right\rangle
\ee
where 
$
\mg=\mk-\mk\ms\mm^{-1}\ms^t\mk
$ depends on the configuration of the network.
We can decompose the stiffness matrix $\mk$ and the structure matrix $\ms$ onto the strong and weak connections, 
\[
\mk=\lp\begin{array}{cc}k\mi_{\rm s} & 0\\0 & k_{\rm w}\mi_{\rm w}\end{array}\rp,
\qquad\ms=\lp\begin{array}{c}\ms_{\rm s}\\\ms_{\rm w}\end{array}\rp.
\]
From the approximation of the dynamic matrix $\mm$ in Eq.(\ref{alpha}), we can decompose it as
\[
\mm=k\sum_{\omega}(\omega^2+\alpha)|\delta{\bf R}_\omega\rangle\langle\delta{\bf R}_\omega|.
\]
Similarly, we write $\mi_{\rm s}$ and $\ms_{\rm s}$ in the same basis and corresponding basis in connection space $|\delta r_\omega\rangle=\frac{1}{\omega}\ms_{\rm s}|\delta{\bf R}_\omega\rangle$,
\[
\ms_{\rm s}=\sum_\omega\omega|\delta r_\omega\rangle\langle\delta{\bf R}_\omega|,\quad
\mi_{\rm s}=\sum_p|\psi_p\rangle\langle\psi_p|+\sum_\omega|\delta r_\omega\rangle\langle\delta r_\omega|
\]
where $|\psi_p\rangle$ defines the null space of the structure $\ms_{\rm s}$ that self-stresses live in. We then get,
\begin{multline}
\label{g0}
G(\Gamma)=\frac{1}{V}\lp k\sum_p|X_p|^2+k\sum_\omega\frac{\alpha}{\omega^2+\alpha}|X_\omega|^2\right.\\
\left.+N_{\rm w}k_{\rm w}a_{\rm w}^2-2k_{\rm w}\sum_\omega\frac{\omega}{\omega^2+\alpha}X_\omega X^{\rm w}_\omega+o(k_{\rm w}^2)\rp,
\end{multline}
where $X_p=\langle\psi_p|\frac{\Delta x\Delta y}{\Delta r}\rangle$, $X_\omega=\langle\delta r_\omega|\frac{\Delta x\Delta y}{\Delta r}\rangle$, $N_{\rm w}a_{\rm w}^2=\sum_{\rm weak}\langle\frac{\Delta x\Delta y}{\Delta r}|\frac{\Delta x\Delta y}{\Delta r}\rangle$, and $X_\omega^{\rm w}=\langle\frac{\Delta x\Delta y}{\Delta r}|\ms_{\rm w}|\delta{\bf R}_\omega\rangle$. 

Finally, we average over the configurations. For the isotropic disordered networks we are dealing with, $\frac{\Delta x\Delta y}{\Delta r}$ should be a random variable distributed evenly around zero independent of the choices of basis. $X_p$, $X_\omega$ are thus sums of $N_{\rm s}$ random variables with zero mean. Central Limit Theorem thus gives,
\be
G=\left\{\begin{array}{cc}
\rho n_cka^2\lp \frac{n}{n_c}-1+\int\rd\omega D(\omega)\frac{\alpha}{\omega^2+\alpha}+\alpha\frac{a_{\rm w}^2}{2a^2}\rp & n>n_c\\
\rho n_cka^2\lp \frac{n}{n_c}\int\rd\omega D(\omega)\frac{\alpha}{\omega^2+\alpha}+\alpha\frac{a_{\rm w}^2}{2a^2}\rp & n<n_c 
\end{array}\right.
\ee
where $\rho=N/V$, $a^2$ is the variance of $X_p$ and $X_\omega$, and $D(\omega)$ is normalized density of vibrational states. For perturbative $\alpha\ll1$, the shear modulus $G\propto n-n_c$ for $n>n_c$, and $G\sim\alpha$ when $n<n_c$. 

}

\subsection{F. Segregation in $A_xB_{1-x}$}
Consider chemical compound $A_xB_{1-x}$, where both $A$ and $B$ atoms, as isotropic particles, possess $d$ degrees of freedom. The number of constraints counting both bond stretching and bending per $B$ satisfies {$n^{B}<n_c=d$} and the number per $A$ {$n^{A}>n_c$}, so that both floppy and rigid networks can be produced by composition. In the range of segregation, there appear a stressed rigid phase with volume fraction $V_r$, concentrations of $B$ $\rho_r^{B}$ and $A$ $\rho_r^{A}$, and a floppy phase with $V_f$, $\rho_f^{B}$ and $\rho_f^{A}$. 

Similar to the counting approximation Eq.(4), vibrational entropy obeys, 
{
\be
\frac{S_{\rm vib}}{N}=V_f[(n_c-n^{B})\rho_f^{B}+(n_c-n^{A})\rho_f^{A}]{\Lambda},
\label{eq_ABsv}
\ee
}
with $\Lambda$ the vibrational entropy gain from each floppy mode. The configurational entropy of two segregated regions is, 
\be
\frac{S_{\rm conf}}{N}=-V_f(\rho_f^{B}\ln\rho_f^{B}+\rho_f^{A}\ln\rho_f^{A})-V_r(\rho_r^{B}\ln\rho_r^{B}+\rho_r^{A}\ln\rho_r^{A}).
\label{eq_ABsc}
\ee

Optimizing entropy with the following constraints,
$
V_f+V_r=1$, 
$
V_f\rho_f^{A}+V_r\rho_r^{A}=x$, 
and
$
V_f\rho_f^{B}+V_r\rho_r^{B}=1-x$, 
we end up with following phase boundaries,
{
\begin{subequations}
\begin{align}
&\rho_f^{A}=\frac{e^{(n_c-n^{B}){\Lambda}}-1}{e^{(n^{A}-n^{B}){\Lambda}}-1};\\
&\rho_r^{A}=\rho_f^{A}e^{(n^{A}-n_c){\Lambda}}=\frac{e^{(n^{A}-n^{B}){\Lambda}}-e^{(n^{A}-n_c){\Lambda}}}{e^{(n^{A}-n^{B}){\Lambda}}-1};\\
&\rho_f^{B}=1-\rho_f^{A};\quad \rho_r^{B}=\rho_f^{B}e^{(n^{B}-n_c){\Lambda}}=1-\rho_r^{A};\\
&V_r=\frac{x-\rho_f^{A}}{\rho_r^{A}-\rho_f^{A}}. 
\end{align}
\label{eq_ABsol}
\end{subequations}
}
The boundary of the heterogeneous phase when self-stress is prohibited is determined by, 
{
\be 
\Lambda=\frac{1}{d}\lp\rho_c^A\ln\frac{\rho_c^A}{{\rho_f^A}}+(1-\rho_c^A)\ln\frac{1-\rho_c^A}{1-{\rho_f^A}}\rp
=\frac{1}{n_c}D(\rho_c|\rho_f).
\label{eq_RPrho}
\ee }
As many constraints are associated with a high valence atom, the configurational entropy cost to generate phase separation is lower than in the network model {by a factor of $n_m$}. So the transition boundary Eq.(\ref{eq_RPrho}) is at a much lower value than Eq.(7), and the segregation happens much easier. 

{
\subsection{G. General interactions.}
We generalize our results to elastic networks of dispersed interactions, still assuming the separation of energy scales. Each pair of neighboring particles either interact through a bond of strength $\kappa$ from some distribution $\rho(\kappa)$ or do not interact, 
\be
P(\kappa)=(1-p)\delta(\kappa)+p\rho(\kappa).
\label{pk}
\ee
where $\delta$ is Dirac delta function. When phases separate, we have a rigid phase of volume $V_r$ with connections characterized by a distribution $P_r(\kappa)$ and a floppy phase of volume $V_f$ and distribution $P_f(\kappa)$. They are constrained by
\begin{subequations}
\begin{align}
& V_f+V_r=1\\
& V_f P_f(\kappa)+V_rP_r(\kappa)=P(\kappa)\\
& \int\rd \kappa P_f(\kappa)=\int\rd \kappa P_r(\kappa) = 1.
\label{eqnorm}
\end{align}
\end{subequations}

The configuration entropy of the layout is,
\begin{multline}
\frac{S_{\rm conf}}{N}=-n_m\lp V_f \int\rd \kappa P_f(\kappa)\ln P_f(\kappa)\right.\\
\left.+V_r \int\rd \kappa P_r(\kappa)\ln P_r(\kappa)\rp
\end{multline}
Without loss of generality, we consider 
the vibrational entropy following
\begin{multline}
\frac{S_{\rm vib}}{N}=V_f\Lambda\int\rd k\eta_f(\kappa)P_f(\kappa)+V_r\Lambda\int\rd \kappa\eta_r(\kappa)P_r(\kappa)\\
=V_f\Lambda\int\rd \kappa\eta(\kappa)P_f(\kappa)+s_0,
\end{multline}
where $\eta(\kappa)=\eta_f(\kappa)-\eta_r(\kappa)$ and $s_0=\Lambda\int\rd \kappa\eta_r(\kappa)P(\kappa)$.  
This linear assumption, however, may just be approximately true, especially when the segregation of weak interactions appear, which is accompanied with a diverging density of soft vibrations~\cite{DeGiuli14b} and contributes nonlinearly. 
We define a marginal network $P_c(\kappa)$ by where the vibrational entropy equals to zero,
$
\int\rd \kappa\eta(\kappa)P_c(\kappa)=0.
$

All together, the total entropy $S_{\rm vib}+S_{\rm conf}$ is optimized by,
\be
\Lambda\eta(\kappa)=\ln\frac{P_f(\kappa)}{P_r(\kappa)}.
\label{log}
\ee
Multiplying both side of Eq.(\ref{log}) with $P_c(\kappa)$ and integrate over $\kappa$, we find a balance condition,
\be
D(P_c|P_f)=\int\rd \kappa P_c(\kappa)\ln\frac{P_c(\kappa)}{P_f(\kappa)}=D(P_c|P_r),
\ee
the relative entropies~\cite{Mezard09} to the critical distribution of the distributions in the rigid and floppy phases are equal. Similarly, we have,
\be
\Lambda\int\rd \kappa\eta(\kappa)P_f(\kappa)=\int\rd \kappa P_f(\kappa)\ln\frac{P_f(\kappa)}{P_r(\kappa)}=D(P_f|Pr),
\label{entr}
\ee
entropic gain per unit volume in floppy phase compensates the relative entropy from the rigid phase to the floppy one.

The general results apply to specific cases. In the network of a single type strong interaction, we have $P(\kappa)=\frac{n_m-n}{n_m}\delta(\kappa)+\frac{n}{n_m}\delta(\kappa-k)$ and $\eta(\kappa)=\frac{n_c}{n_m}\delta(\kappa)-\frac{n_m-n_c}{n_m}\delta(\kappa-k)$. In the network of compounds $A_xB_{1-x}$, $\kappa$ labels different chemical elements, $P(\kappa)=\rho^A\delta_{\kappa,A}+\rho^B\delta_{\kappa,B}$ and $\eta(\kappa)=(n_c-n^B)\delta_{\kappa,B}+(n_c-n^A)\delta_{\kappa,A}$, where $\delta$ is Kronecker delta symbol.

\begin{figure}[htbp]
\centering
\includegraphics[width=.9\columnwidth]{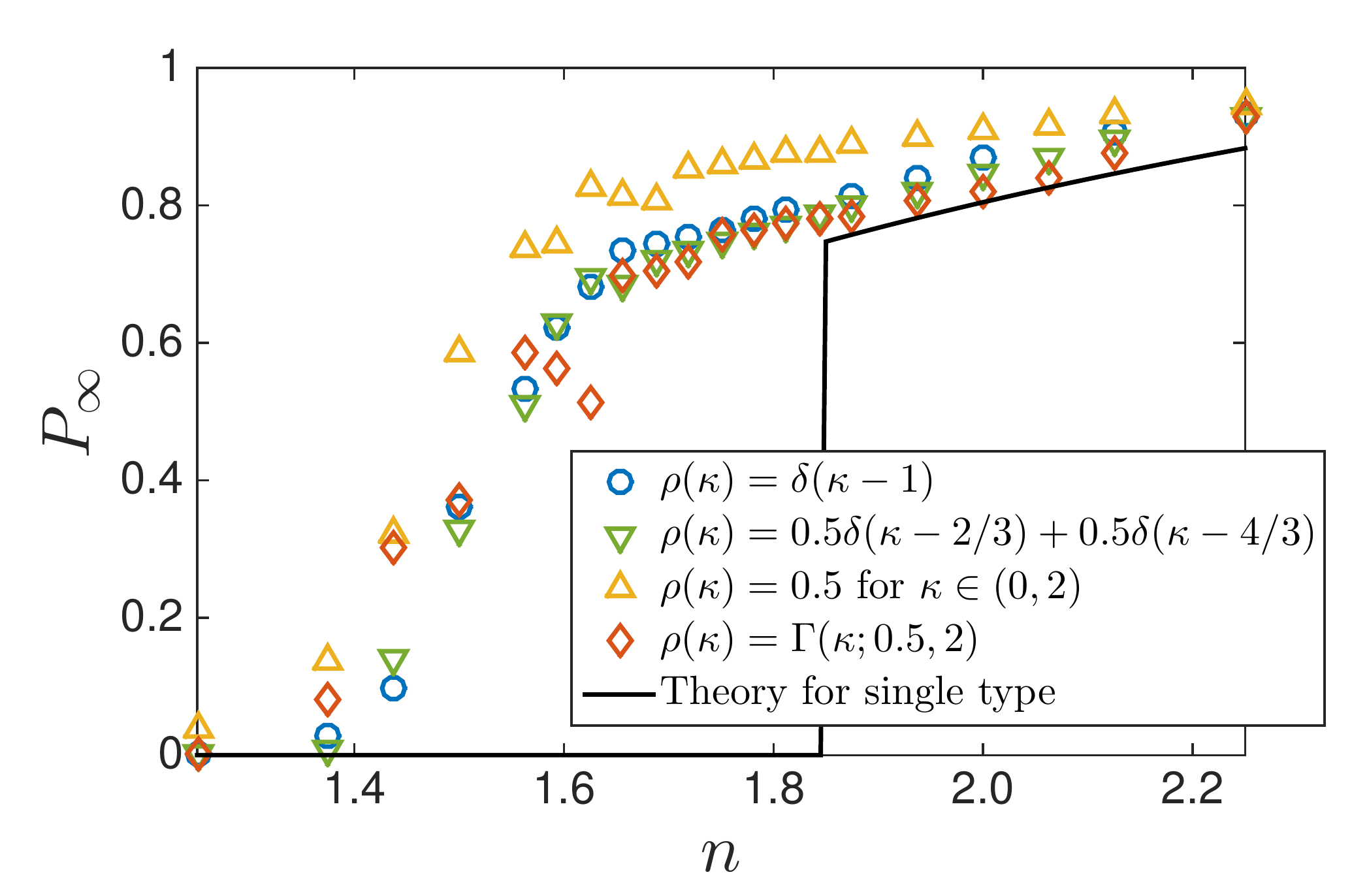}\\
\caption{\small{Probability of the bond in the rigidity percolating cluster $P_\infty$ versus the average number of constraints $n$ for $N=256$. The stiffnesses of the constraints $k$ are drawn from the distributions $\rho(\kappa)$, including single type (blue circles), two types (green triangles), uniform distribution (yellow triangles), and Gamma distribution (red diamonds).}}\label{Pinfd}
\end{figure}

{\bf Numerical evidence of separation.} We confirm numerically the robustness of our prediction on phase separation independent of our choice of single type of strong interactions. We have considered the bi-disperse, uniformly distributed, and Gamma distributed interaction strengths. As shown in Fig.~\ref{Pinfd}, independent of the choice of the distributions, the rigidity consistently percolates below the Maxwell point $n_c=2$, because of the existence of the highly-connected rigid phase resulted from the phase separation. 

}

\subsection{H. The nonlinear limit}

In the main text, we have focused on the thermal vibrations in the linear range in Eq.(2), valid in the low temperature limit. In order to see when the conclusions are valid and how entropy directs the network organization in the high temperature limit, we consider the nonlinear responses acting as a cutoff, than which the range of the linear vibration $T/\omega^2$ can not be larger. 
It's reasonable to assume that the nonlinear response starts to effect when the relative displacement of atoms is larger than the Lindemann's criterion~\cite{Lindemann10}, about 0.15 time of the typical atom distance. 
\be
\label{e1}
S_{\rm vib}(\Gamma)=\sum_{\omega}\min[-\ln\omega+\frac{1}{2}\ln T, -\frac{1}{d}\ln P(\omega)+c]
\ee
where $P(\omega)=\sum_i\delta{\bf R}_i(\omega)^4$ is the participation ratio of the corresponding eigenmode $\delta{\bf R}_i(\omega)$, which estimates the number of atoms involved in given mode. So $ P(\omega)^{-1/d}$ gives the relative displacement of two neighbors in the unit of Lindemann's distance. $c$ is the constant determined by the range of nonlinear response. 

\begin{figure}[h!]
\centering
\includegraphics[width=1.0\columnwidth]{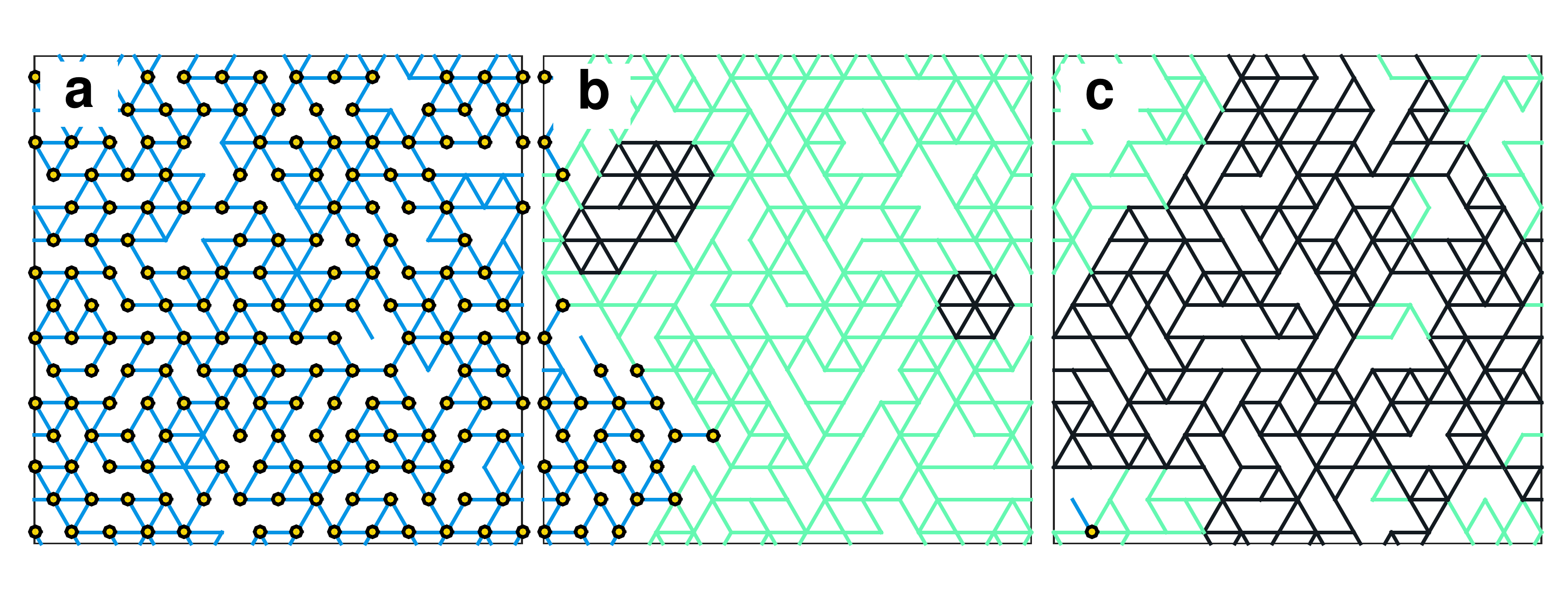}\\
\caption{\small{Typical network structures (a) below, (b) at, and (c) above the Maxwell point $z_c$ obtained with vibrational entropy at high temperature limit Eq.(\ref{e3}).}}\label{tinf}
\end{figure}

The linear limit Eq.(2) breaks down at the high temperate $\ln T\gtrsim 2c$, where vibrational phase space is cutoff by the nonlinear response in each degrees of freedom,
\be
S_{\rm vib}=-\frac{1}{d}\sum_{\omega}\ln P(\omega).
\label{e3}
\ee

The typical structures maximizing Eq.(\ref{e3}) are shown in Fig.~\ref{tinf}. In contrast to the heterogeneous effect of vibrational entropy discussed in the main text, it improves the homogeneity of the network structures, and the rigidity of the resulted networks again converges to the scenario discussed in mean-field theory with a sharp jump $P_\infty$ at {$n_c$}. 

\end{appendix}
\end{document}